\documentclass[aps,pre,twocolumn,superscriptaddress,longbibliography]{revtex4-1}
\usepackage{color}
\usepackage[usenames,dvipsnames]{xcolor}

\usepackage{graphicx}
\usepackage{epsfig}
\usepackage{dcolumn}
\usepackage{bm}
\usepackage{mathrsfs}
\usepackage{multirow}
\usepackage[all]{xy}
\usepackage{pbox}
\usepackage{amssymb}
\usepackage{amsmath}
\usepackage{amsfonts}

\usepackage{color}
\definecolor{myblue}{RGB}{65,105,225}
\definecolor{mygreen}{RGB}{34,139,34}
\definecolor{myorange}{RGB}{255,69,0}
\usepackage[colorlinks,linkcolor=myorange,urlcolor=myblue,citecolor=myblue]{hyperref}

\def\(({\left(}
\def\)){\right)}
\def\[[{\left[}
\def\]]{\right]}

\newcommand{\beq}{\begin{equation}}
\newcommand{\eeq}{\end{equation}}

\newcommand{\ben}{\begin{eqnarray}}
\newcommand{\een}{\end{eqnarray}}

\newcommand{\la}{\langle}
\newcommand{\ra}{\rangle}
\newcommand{\be}{\begin{equation}}
\newcommand{\ee}{\end{equation}}

\newcommand{\vE}{{\bf{E}}}
\newcommand{\vq}{{\bf{q}}}

\newcommand{\vlamb}{{\bm{\lambda}}}

\newcommand{\vz}{{\bf{z}}}
\newcommand{\mA}{{\hat{\cal A}}}
\newcommand{\hD}{{\hat{D}}}
\newcommand{\hS}{{\hat{\sigma}}}

\begin{document}  

\title{Sampling rare events across dynamical phase transitions}

\author{Carlos P\'erez-Espigares}
\affiliation{Departamento de Electromagnetismo y F\'{\i}sica de la Materia, and Institute Carlos I for Theoretical and Computational Physics, Universidad de Granada, Granada 18071, Spain}

\author{Pablo I. Hurtado}
\affiliation{Departamento de Electromagnetismo y F\'{\i}sica de la Materia, and Institute Carlos I for Theoretical and Computational Physics, Universidad de Granada, Granada 18071, Spain}

\date{\today}

\begin{abstract}
Interacting particle systems with many degrees of freedom may undergo phase transitions to sustain atypical fluctuations of dynamical observables such as the current or the activity. This leads in some cases to symmetry-broken space-time trajectories which enhance the probability of such events due to the emergence of ordered structures. Despite their conceptual and practical importance, these dynamical phase transitions (DPTs) at the trajectory level are difficult to characterize due to the low probability of their occurrence. However, during the last decade advanced computational techniques have been developed to measure rare events in simulations of many-particle systems that allow for the first time the direct observation and characterization of these DPTs. Here we review the application of a particular rare-event simulation technique, based on cloning Monte Carlo methods, to characterize DPTs in paradigmatic stochastic lattice gases. In particular, we describe in detail some tricks and tips of the trade, paying special attention to the measurement of order parameters capturing the physics of the different DPTs, as well as to the finite-size effects (both in the system size and number of clones) that affect the measurements. Overall, we provide a consistent picture of the phenomenology associated with DPTs and their measurement.
\end{abstract}


\maketitle 

{\bf{Large dynamical fluctuations are realizations of the dynamics sustained during a long period of time which deviate very much from their average value. Despite being very unlikely to occur, these fluctuations appear in many different systems carrying a large impact. Examples range from oceanic rogue waves, or chemical reaction kinetics to climate changes or stock market crashes. In the context of nonequilibrium systems, where we cannot derive the macroscopic properties from the  Boltzmann-Gibbs distribution, the study of these fluctuations has led to important advances such as the fluctuation theorems and the formulation of a macroscopic fluctuation theory for driven diffusive systems. However, one of the main challenges to make further progress is precisely that large fluctuations are difficult to be observed as their probability is extremely low. Thus, much effort is currently devoted to implement efficient algorithms allowing for the measurement of such atypical events. In this paper we review the application of a computational technique, based on population dynamics, to measure large fluctuations of the current sustained in driven diffusive systems. We focus our attention on the application of the algorithm to the study of the so-called dynamical phase transitions, which appear as a change in the trajectories of the system in order to maximize the probability of sustaining a large fluctuation. Transitions of this kind are unveiled for several paradigmatic stochastic lattice gases after introducing the correct order parameters.}}

\section{Introduction}
\label{s0}

Phase transitions appear ubiquitously in nature, from cosmological scales to the quantum world of elementary particles. Consequently, the theory of critical phenomena has become one of the cornerstones of modern theoretical physics \cite{binney92a,zinn-justin02a}. According to the modern classification  (which is based on the classical Ehrenfest scheme)  
most phase transitions can be broadly divided into two categories: first-order or discontinuous, and second-order or continuous \cite{binney92a,zinn-justin02a}, although other types do exist which challenge this classification.
In a typical second-order phase transition, some type of order emerges \emph{continuously} at a critical point of a control parameter. This peculiar change is captured by a so-called \emph{order parameter}, and typically signals the spontaneous breaking of a symmetry and an associated non-analyticity of the relevant thermodynamic potential. 
Conversely, first-order transitions are characterized by an \emph{abrupt} jump in the order parameter related to a kink in the thermodynamic potential, which leads to a coexistence between different phases \cite{binney92a,zinn-justin02a}. 

In recent years these ideas have been generalized to the realm of fluctuations, where \emph{dynamical phase transitions} 
have been identified in many different systems \cite{bertini05a,bodineau05a,harris05a,bertini06a,bodineau07a,lecomte07b,lecomte07c,garrahan07a,bodineau08a,garrahan09a,hedges09a,chandler10a,garrahan10a,hurtado11a,garrahan11a,pitard11a,genway12a,ates12a,speck12a,perez-espigares13a,harris13a,villavicencio14a,lesanovsky13a,hurtado14a,vaikuntanathan14a,manzano14a,jack15a,shpielberg16a,zarfaty16a,tsobgni16a,manzano16a,lazarescu17a,brandner17a,karevski17a,carollo17a,baek17a,tizon-escamilla17b,shpielberg17a,pinchaipat17a,abou18a,manzano18a,baek18a,shpielberg18a,perez-espigares18b,perez-espigares18c,chleboun18a,klymko18a,whitelam18a,vroylandt18a,rotondo18a,buca19a,doyon19a}.
But what is \emph{dynamical} about dynamical phase transitions? In contrast to standard critical phenomena, which occur at the configurational level when varying a control parameter such as temperature or magnetic field,
DPTs appear in trajectory space when conditioning a system of interest to sustain an unlikely value of dynamical observables such as the time-integrated current or the activity (which are key magnitudes when studying respectively nonequilibrium systems and amorphous solids).
The different dynamical phases that appear correspond to different types of trajectories adopted by the system during these rare events. 
Interestingly, some dynamical phases and their corresponding trajectories turn out to be 
far more probable than anticipated due to the emergence of ordered structures such as traveling waves \cite{bodineau05a,hurtado11a,perez-espigares13a,karevski17a}, condensates \cite{harris05a,harris13a,chleboun18a} or hyperuniform states \cite{jack15a,carollo17a,carollo18a}.
Another hallmark of a DPT is the appearance of non-analyticities and Lee-Yang singularities \cite{yang52a,arndt00a,blythe02a,dammer02a,blythe03a,flindt13a,hickey14a,brandner17a} in the so-called large deviation function (LDF) which controls the probability of fluctuations. 
This is a finding of crucial importance particularly in nonequilibrium physics, as these LDFs play a role akin to the equilibrium thermodynamic potentials for nonequilibrium systems, where no bottom-up approach exists yet connecting microscopic dynamics with macroscopic properties \cite{bertini15a,derrida07a,barre18a}. 

Symmetry-breaking DPTs are particularly interesting, and their analogy with standard critical phenomena is intriguing. 
In a standard second-order critical point, a (continuous or discrete) symmetry of the governing action is eventually broken, meaning that the system ground state beyond the critical point has
less symmetries than the original action. Symmetry is however recovered by the appearance of different (symmetry-broken) ground states, which map onto each other under the symmetry operator. Remarkably, a similar picture arises in DPTs, but this time at the level of trajectories. To better understand this point, note that the probability of observing a given long-time fluctuation of a dynamical observable is dominated by the probability of the most probable trajectory (or \emph{optimal path}) leading to such fluctuation \cite{hurtado14a,bertini15a}. This defines a sort of dynamical \emph{ground state} for each fluctuation, i.e.~its optimal path, whose properties are crucial to shed light on the physics of the problem of interest \cite{perez-espigares15a,perez-espigares16a,tizon-escamilla17a}.  
The action-like functional 
describing the statistical weight of paths in phase space may have some symmetries (as e.g.~time-translation invariance, particle-hole exchange, etc.) which are typically inherited by the associated optimal paths. However, at a second-order-like DPT, the symmetry of the trajectory action is broken: optimal paths do not share the symmetry of their action, but
symmetry is restored by the appearance of degenerate optimal trajectories linked by the symmetry transformation. This analogy can be further exploited to obtain deep insights into symmetry-breaking DPTs.

In addition to their conceptual importance, DPTs play also a key role to understand the physics of different systems, from glass formers \cite{garrahan07a,garrahan09a,hedges09a,chandler10a,pitard11a,speck12a,pinchaipat17a,abou18a} to superconducting transistors and micromasers \cite{garrahan11a,genway12a}. There have been also recent applications of DPTs to design quantum thermal switches \cite{manzano14a,manzano16a,manzano18a}, i.e. quantum devices where the heat current flowing between hot and cold reservoirs can be completely blocked, modulated or turned on at will.
Furthermore, another interesting possibility opens up by noting that rare events can be turned to typical with the use of Doob's h-transform \cite{doob57a,jack10a,chetrite15a,chetrite15b} or external fields with optimal dissipation \cite{bertini15a}. This can be then used to exploit existing DPTs to engineer and control complex systems with a desired statistics \emph{on demand} \cite{carollo18b}, a possibility which is being currently explored. 

Despite their relevance, observing and characterizing DPTs is a challenging task, the reason being that the spontaneous emergence of large (rare) fluctuations in many-body systems is generally unlikely. However, during the last few years, two new powerful and general methods have appeared to investigate fluctuating behavior in many-particle systems that are broadening our understanding of DPTs. On one hand, at the theoretical level, a macroscopic fluctuation theory (MFT) has been formulated \cite{bertini15a} which offers variational equations to rationalize dynamical fluctuations and the associated LDFs in interacting many-particle systems arbitrarily far from equilibrium, starting from their fluctuating hydrodynamic description and a few transport coefficients. This deep and rich theoretical scheme won't be the focus of this paper, though we will refer to some of its predictions; reviews on this formalism can be found elsewhere \cite{bertini15a,derrida07a,hurtado14a}. A second tool which has reinvigorated the study of DPTs is the development of advanced computational methods to directly measure LDFs and the associated optimal paths in simulations of interacting many-particle systems \cite{giardina06a,lecomte07a,tailleur09a,giardina11a}. These numerical methods amount to modifying the system dynamics so that the rare events responsible for a large deviation become no longer rare, and involve the simultaneous evolution of multiple copies or \emph{clones} of the system of interest, which replicate or die in time according to their statistical weight, a technique based on the Diffusion Monte Carlo method of quantum mechanics \cite{anderson75a}. The application of these new tools to simple models, particularly stochastic lattice gases, is providing intriguing evidences of the existence of rich and fundamental structures in the fluctuating behavior of nonequilibrium systems, which emerge mainly via DPTs, crucial to crack this long-unsolved problem.

The aim of this paper consists in reviewing the application of the cloning Monte Carlo method to understand the physics behind a number of dynamical phase transitions of theoretical interest. In particular, we will describe in detail some tricks of the trade, with an emphasis on the definition and measurement of order parameters capturing the physics of the different DPTs, the characterization of the optimal paths responsible for a fluctuation, as well as the finite-size effects (both in the system size and number of clones) that affect the measurements. In order to do so, we first provide a brief description of the cloning Monte Carlo method in Section \ref{s2}, both in its discrete- and continuous-time versions, together with an analysis of the effect of the finite number of clones on the large deviation function estimators. Once the main computational tools have been introduced, we set out to describe some intriguing results for DPTs in the current statistics of two paradigmatic models of transport, namely the Kipnis-Marchioro-Presutti (KMP) model of heat transport \cite{kipnis82a} and the weakly asymmetric simple exclusion process (WASEP) \cite{derrida98a}. In Section \ref{s3} we study the spontaneous breaking of time-translation symmetry at the trajectory level in periodic systems. We do so 
in the $1d$ KMP model, where a DPT into a dynamical phase dominated by ballistic energy packets has been found, and in $1d$ WASEP, where jammed density-wave trajectories dominate low-current fluctuations. In Section \ref{s4} we explore numerically the important role of dimensionality on dynamical phase transitions by investigating vector current statistics in $2d$ WASEP. Interestingly, the complex interplay among an external field, the possible system anisotropy, and vector currents in $d>1$ leads to a rich phase diagram at the fluctuating level, with different symmetry-broken fluctuation phases separated by lines of first- and second-order DPTs. This remarkable competition between different dynamical phases 
is due to the appearance of a structured vector field coupled to the current, a key feature of high-dimensional (realistic) systems. Section \ref{s5} is devoted to study a different type of symmetry-breaking phenomenon at the trajectory level which appears in open systems, i.e. coupled to boundary reservoirs which may drive the system out of equilibrium by imposing an external gradient (of e.g. density or temperature). In this case the symmetry that is broken at the DPT is the particle-hole symmetry --a $\mathbb{Z}_2$ discrete symmetry--, and we show that the transition persists in the presence of arbitrarily strong (but symmetric) boundary gradients. Finally, Section \ref{s6} discusses the results presented in this review from a general point of view, providing also some outlook on the work that remains to be done in a near future.

\section{The statistical physics of trajectories}
\label{s1}

As mentioned earlier, we will be interested in this paper on DPTs emerging in the statistics of a dynamical observable such as the space\&time-averaged current. However, the following large deviation formalism applies as well to any time-integrated observable. In order not to clutter our notation we particularize our discussion to one-dimensional ($1d$) systems (unless otherwise stated), though extensions to arbitrary dimension and vector currents are straightforward \cite{hurtado14a}. With this aim in mind, we hence consider the statistical physics of an ensemble of trajectories conditioned to a given total current $Q$ integrated over a long time $t$. 
This trajectory ensemble is fully characterized by the probability $P_{t}(Q)$ of all trajectories of duration $t$ with total current $Q$. In most cases of interest, this probability can be shown to obey a \emph{large-deviation principle} for long times $t$ \cite{bertini15a,derrida07a,hurtado14a}, i.e. $P_{t}(Q)$ scales in this limit as
\be
P_{t}(Q) \asymp \exp[+t F(Q/t))] \, , 
\label{ldp}
\ee
where the symbol "$\asymp$" represents asymptotic logarithmic equality, i.e.
\be
\lim_{t\to\infty} \frac{1}{t} \ln P_{t}(Q=\hat{q} t) = F(\hat{q}) \, .
\label{asymp}
\ee
The function $F(\hat{q})$ in Eq. (\ref{ldp}) above, with $\hat{q}=Q/t$, defines the large deviation function (LDF) of the current. This LDF is a measure of the (exponential) rate at which the probability of observing an empirical current $\hat{q}$ --appreciably different from its steady-state value $\la \hat{q}\ra$-- decays as $t$ increases. Note that this implies that $F(\la \hat{q} \ra)=0$. In the spirit of ensemble theory of equilibrium statistical mechanics, one can also characterize the system in terms of a \emph{dynamical partition function} 
\be
Z_{t}(\lambda)=\sum_Q P_{t}(Q) \text{e}^{\lambda Q}\, ,
\label{Zlambdam}
\ee
or equivalently by the associated \emph{dynamical free energy} (dFE) 
\be
\theta(\lambda)=\lim_{t\to\infty} \frac{1}{t} \ln Z_{t}(\lambda) \, ,
\label{dFE}
\ee
which is nothing but the Legendre transform of the current LDF, namely
\be
\theta(\lambda)=\max_{\hat{q}}\left[F(\hat{q}) + \lambda \hat{q} \right] \, .
\label{Legendre}
\ee
The intensive parameter $\lambda$ is conjugated to the time-extensive current $Q$. This relation is equivalent to the connection between temperature and energy in equilibrium systems. However, and unlike temperature, the parameter $\lambda$ is non-physical and cannot be directly manipulated is experiments, a main difficulty when studying DPTs which can be however circumvented using the \emph{active interpretation} of fluctuation formulas \cite{bertini15a}. In any case, fixing $\lambda$ to a constant value is equivalent to conditioning the system of interest to have a time-averaged (intensive) current $\hat{q}_\lambda \equiv Q_\lambda/t = \partial_\lambda \theta(\lambda)$, so by varying $\lambda$ one may change the associated current and therefore move from one dynamical phase to another. 

Most of the systems whose dynamical fluctuations we are interested in can be described at the mesoscopic level by a locally-conserved density field $\rho(x,t)$ which evolves in time according to a fluctuating hydrodynamic equation. This can be seen as a continuity equation $\partial_t \rho + \partial_x j =0$ coupling the local density field $\rho(x,t)$ with a fluctuating local current $j(x,t)$. This current field typically obeys Fick's (or Fourier's) law, and includes a noise term which captures all the fast degrees of freedom which have been re-summed in the coarse-graining procedure leading to this mesoscopic description, namely
\be
j(x,t)=-D(\rho)\partial_x \rho(x,t) + \sigma(\rho) E + \xi(x,t) \, .
\label{current}
\ee
Here $D(\rho)$ and $\sigma(\rho)$ are the diffusivity and mobility transport coefficients, respectively, and $E$ is a possible external field applied on the system of interest. The stochastic field $\xi$ is a Gaussian white noise, with $\la\xi\ra=0$ and 
$\la \xi(x,t)\xi(x',t')\ra=L^{-1} \sigma(\rho) \delta(x-x') \delta(t-t')$, where $L$ is the size of the system.
Moreover, this fluctuating hydrodynamic description must be supplemented by appropriate boundary conditions, which can be either periodic or open (see below). Interestingly, we can associate with any trajectory $\{\rho(x,t),j(x,t)\}_0^{\tau}$ of duration $\tau$ in mesoscopic phase space an empirical space\&time-averaged current $q=\tau^{-1} \int_0^{\tau} dt \int_0^1 dx~j(x,t)$. Due to the diffusive scaling when going from the microscopic description to a mesoscopic stochastic field theory, the relation between the macroscopic current $q$ and the microscopic one $\hat{q}$ can be shown to be $q=L\hat{q}$ \cite{derrida07a}. The probability $P(\{\rho,j\}_0^{\tau})$ of any trajectory can be computed using a path integral formalism \cite{hurtado14a,bertini15a,derrida07a}, and scales in the large-size limit
as $P(\{\rho,j\}_0^{\tau})\asymp \exp\{-L\, {\cal I}_{\tau}[\rho,j]\}$, with an action \cite{bertini15a}
\be
{\cal I}_{\tau}[\rho,j] = \int_0^{\tau} dt \int_0^1 dx \frac{\displaystyle \Big(j+D(\rho)\partial_x\rho - \sigma(\rho) E \Big)^2}{\displaystyle 2\sigma(\rho)} \, .
\label{MFT1}
\ee
This probability measure represents the ensemble of space-time trajectories at this mesoscopic level of description. The probability of a given current $q$ can be now obtained by minimizing the action functional (\ref{MFT1}) over all trajectories sustaining such current, leading in the long-time limit to $P(q)\asymp e^{\tau L G(q)}$, with a current LDF
\be
G(q)=-\lim_{\tau \rightarrow \infty} \frac{1}{\tau}\min_{\{\rho,j\}_0^{\tau}}{\hspace{-0.15cm}}^* ~{\cal I}_{\tau} (\rho,j) \, ,
\label{qLDF}
\ee
where $^*$ means that the minimization procedure must be compatible with the prescribed constraints ($q$, boundary conditions, $\partial_t \rho + \partial_x j =0$, etc). The optimal trajectories $\rho_q(x,t)$ and $j_q(x,t)$ solution of this variational problem define the path the system follows to sustain a current $q$ over a long period of time, and turn out to be time-independent in many cases (a conjecture known as {\em additivity principle} \cite{bodineau04a}). In this case, the current LDF simplifies to
\be
G(q)= -\min_{\rho(x)} \int_0^1 dx~\frac{\displaystyle \left[q + D(\rho)\partial_x \rho -\sigma(\rho) E \right]^2}{\displaystyle 2\sigma(\rho)} \, .
\label{LDFap}
\ee
As we will analyze below, sometimes this time-independent solution becomes unstable for large enough current fluctuations under periodic boundary conditions, a DPT leading to a more complex (and now time-dependent) traveling wave optimal trajectory $\rho_q(x,t)=\omega_q(x-vt)$ characterized by a non-trivial velocity $v$. 
More details on this MFT problem can be found elsewhere \cite{bodineau05a,hurtado11a,perez-espigares13a,hurtado14a}.

Note that both in the general case and in the time-independent approximation, equivalent variational problems can be formulated for the dynamical free energy $\mu(\lambda)$ \cite{bertini15a,derrida07a,hurtado14a}, which reads
\be
\mu(\lambda)=\max_{q}\left[G(q) + \lambda q \right] \, .
\label{LegendreMacro}
\ee
The relation between the microscopic and macroscopic LDFs and their corresponding dynamical free energies are $F(\hat{q})=L^{-1}G(q)$ and $\theta(\lambda)=L^{-1}\mu(\lambda)$ in $d=1$ \cite{hurtado14a,derrida07a}. Dynamical phase transitions correspond to singularities in the dFE $\mu(\lambda)$ or equivalently in the current LDF $G(q)$. These singularities (which typically are first- or second-order, as explained in the previous section) are accompanied by peculiar changes in the most probable trajectories or optimal paths associated with these fluctuations. For instance, a broad current interval may be dominated by time-independent optimal trajectories, i.e.~by the additivity principle solution \cite{bodineau04a,hurtado09c}, while in some cases it is known that this additivity conjecture breaks down at some critical current, beyond which time-dependent traveling-wave-like optimal paths dominate the variational problem. This singular change can be rationalized as a second-order DPT where time-translation symmetry is broken \cite{bodineau05a,bertini05a,bertini06a,hurtado11a,perez-espigares13a,hurtado14a}. Similarly, systems with particle-hole symmetry may exhibit regimes of current fluctuations where the dominant trajectory (or dynamical ground state as termed in the introduction) breaks such particle-hole invariance \cite{baek17a,baek18a,perez-espigares18c}.

\section{Sampling rare events with cloning Monte Carlo}
\label{s2}

We now turn our attention to the measurement and characterization of rare events in simulations of stochastic many-particle systems. This is of course a difficult task as these rare events 
are exponentially unlikely (as measured by their associated large deviation function), and hence typically hard to observe in experiments or simulations. Different computational strategies have been proposed to solve this issue, ranging from transition path sampling techniques \cite{dellago02a,bolhuis02a} or density matrix renormalization group methods \cite{gorissen09a,gorissen12a,gorissen12b} to cloning Monte Carlo simulations \cite{giardina06a,lecomte07a,tailleur09a,giardina11a}. In this paper we will focus on the latter method, which is best suited to measure in nonequilibrium situations the probability of a large deviation for time-extensive observables such as the current or the activity. We will describe two different versions of this general method, namely as applied to stochastic many-particle systems evolving with discrete- and continuous-time dynamics. 
Computing the probability of a rare event is equivalent to calculating the leading spectral properties of a \emph{tilted} or \emph{deformed} stochastic matrix which no longer conserves probability. In this way, the key idea consists in reading this tilted matrix as a new dynamical generator which makes the rare event typical, interpreting the non-conservation of probability as an effective population dynamics where different clones or copies of the system reproduce and die according to their statistical weights in the tilted dynamics.

Before proceeding, we note that the material in this section has been described in more detail elsewhere \cite{giardina06a,lecomte07a,tailleur09a,hurtado09a,hurtado10a,giardina11a,hurtado14a}, here we summarize the main ideas and results for clarity and completeness purposes.

\subsection{Discrete-time cloning algorithm}
\label{s2a}

We hence consider a stochastic many-particle Markovian system in discrete time whose microscopic configuration at time $t$ is given by $C_t$. We will be here interested in the statistics of trajectories, which are nothing but sequences of configurations $\{C_0, C_1,\ldots, C_t\}$.
If $W_{C'C}$ is the transition probability from configuration $C$ to configuration $C'$ in the stochastic model of interest, then by virtue of the Markov property the probability of a path is given by
\be
\mathbb{W}_{C_0, C_1,\ldots, C_t} = W_{C_t C_{t-1}}\ldots W_{C_1 C_0} P(C_0) \, ,
\label{ppath}
\ee
where $P(C_0)$ is an initial distribution for the first configuration. Associated with each jump in configuration space, $C \to C'$, one can define the elementary current $q_{C' C}$ involved in this microscopic transition.
The probability of observing a total time-integrated current $Q$ after a time $t$ 
is now given by the sum of the probabilities of all paths compatible with such current, and hence can be formally written as
\be
P_t(Q) = \sum_{C_{t}..C_0} W_{C_t C_{t-1}}..W_{C_1 C_0} P(C_0) \, \delta \left(Q - \sum_{k=0}^{t-1} q_{C_{k+1} C_k}\right) \, ,
\label{recurr1}
\ee
where the Dirac delta-function implements the constraint on the total current. 
Working with a global constraint on the total current as in Eq. (\ref{recurr1}) is typically difficult, so it seems convenient to \emph{change ensemble} in the spirit of equilibrium statistical mechanics and work with the moment-generating function of the current distribution, i.e. the dynamical partition function 
$Z_t(\lambda)=\sum_Q P_t(Q) e^{\lambda Q}$ of previous section, see Eq. (\ref{Zlambdam}). Using (\ref{recurr1}) in the definition of $Z_t(\lambda)$, we find that
\be 
Z_t(\lambda) = \sum_{C_{t}..C_0} W_{C_t C_{t-1}}..W_{C_1 C_0} P(C_0) \, \text{e}^{\lambda \sum_{k=0}^{t-1} q_{C_{k+1} C_k}} \, .  
\label{pi1}
\ee
This expression suggests defining a modified dynamics 
\be
\tilde{W}_{C' C}\equiv \text{e}^{\lambda q_{C' C}}\, W_{C' C} \, ,
\label{moddyn}
\ee
so the dynamical partition function can be now written as
\begin{equation}
Z_t(\lambda)  = \sum_{C_{t}\ldots C_0} \tilde{W}_{C_t C_{t-1}} \ldots \tilde{W}_{C_1 C_0} P(C_0)  \, .
\label{pi2}
\end{equation}
However, the transition matrix with elements $\tilde{W}_{C' C}$ does not define a proper stochastic matrix, as the modified dynamics is not normalized, $\sum_{C'} \tilde{W}_{C' C}\neq 1$. Introducing now the exit rates 
\be
Y_C \equiv \sum_{C'} \tilde{W}_{C' C} \, ,
\label{Yc}
\ee
we can define now a \emph{bona fide} normalized dynamics
\be
W'_{C' C}\equiv \frac{\displaystyle 1}{\displaystyle Y_C} \tilde{W}_{C' C} = \frac{\displaystyle \text{e}^{\lambda q_{C' C}}}{\displaystyle Y_C}\, W_{C' C} \, ,
\label{Wmod}
\ee
which is now correctly normalized. The dynamical partition function now reads
\be
Z_t(\lambda)  = \sum_{C_{t}\ldots C_0}  W'_{C_t C_{t-1}} Y_{C_{t-1}} \ldots  W'_{C_1 C_0} Y_{C_0} P(C_0)\, .
\label{pilambda}
\ee

This sum over paths can be read 
from a computational point of view in terms of a population dynamics, which combines
a cloning or replication step proportional to the exit rate of a configuration, $Y_C$, 
followed by a configuration jump governed by the modified normalized dynamics $W'_{C' C}$. The cloning step can be realized by substituting a particular configuration $C$ by an integer number of copies (0, 1, 2, \ldots) with average $Y_C$, which then evolve independently according to the stochastic matrix $W'_{C' C}$.
This population dynamics typically gives rise to an exponential increase or reduction in the number of copies of the system. Indeed, for a given trajectory $\{C_0,\ldots, C_t \}$, the number ${\cal N}(C_0,\ldots, C_t;t)$ of copies 
at time $t$ obeys a simple recurrence relation 
\be
{\cal N}(C_0,\ldots, C_t;t)=W'_{C_t C_{t-1}} Y_{C_{t-1}}  {\cal N}(C_0,\ldots, C_{t-1};t-1) \, ,
\label{recurr2}
\ee
so iterating we arrive at 
\begin{equation}
{\cal N}(C_0,\ldots, C_t;t)= W'_{C_t C_{t-1}} Y_{C_{t-1}} \ldots W'_{C_1 C_0} Y_{C_0} {\cal N}(C_0;0) \, , 
\label{fraccop}
\end{equation}
where ${\cal N}(C_0;0) = N_c P(C_0)$, with $N_c$ the initial total number of copies. The average $\la {\cal N}(t)\ra$ of the total number of clones at time $t$
is obtained by summing over all trajectories of such duration, and is proportional to the dynamical partition function $Z_t(\lambda)$, see Eq. (\ref{pilambda}) above,
\ben
\la {\cal N}(t)\ra &=& \sum_{C_{t}\ldots C_0} {\cal N}(C_0,\ldots, C_t;t) = N_c ~Z_t(\lambda)\nonumber \\
&\asymp&  N_c ~\text{e}^{+t\theta(\lambda)} \, , \label{nclones}
\een
where we have used in the last step the exponential scaling of the dynamical partition function at long times, see Eq. (\ref{dFE}), which demonstrates the exponential scaling of the population of clones as time evolves. In this way, 
\be
Z_t(\lambda)= \frac{\la {\cal N}(t)\ra}{N_c} = \frac{\la {\cal N}(t)\ra}{\la {\cal N}(t-1)\ra} \frac{\la {\cal N}(t-1)\ra}{\la {\cal N}(t-2)\ra} \ldots \frac{\la {\cal N}(1)\ra}{N_c} \, .
\ee 
The last equality trivially splits the total growth of the population into the different increments at each time step. This allows us to deal numerically with the exponential explosion (or implosion) of the total population. Indeed, we find for the dynamical free energy, see Eq. (\ref{dFE}),
\be
\theta(\lambda) =\lim_{t\to\infty} \frac{1}{t} \sum_{k=1}^t \ln \left( \frac{\la {\cal N}(k)\ra}{\la {\cal N}(k-1)\ra} \right) \, ,
\ee
with the definition $\la {\cal N}(0)\ra\equiv N_c$.

\begin{figure}
\centerline{
\includegraphics[width=8cm]{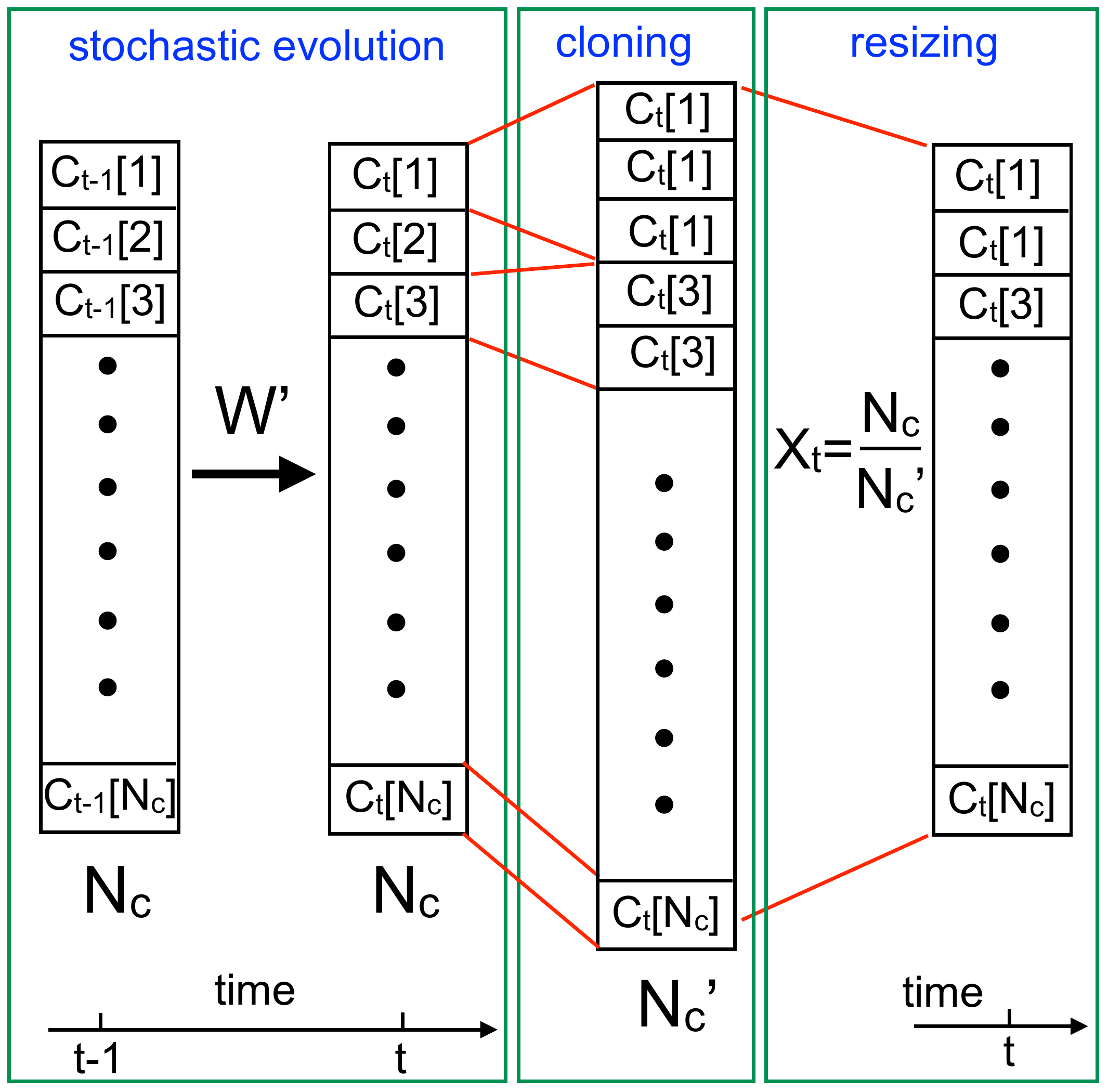}
}
\caption{Sketch of the cloning algorithm in discrete time during the evaluation of the large deviation function.}
\label{sketchclon}
\end{figure}

We are now in position to exploit the previous ideas to sample the sum over paths that define the dynamical partition function $Z_t(\lambda)$
using Monte Carlo techniques, in particular inspired by the Diffusion Monte Carlo method of quantum mechanics \cite{anderson75a}. This sampling can be realized by considering a large ensemble of $N_c\gg 1$ copies of the system of interest (also dubbed \emph{clones}, and hence the name of cloning Monte Carlo methods) which evolve sequentially according to the following steps \cite{giardina06a,giardina11a}:
\begin{enumerate}
\item[(a)] Each copy evolves independently as dictated by the modified normalized dynamics $W'_{C' C}$, which favors certain local jumps depending on the current $q_{C' C}$ involved, see Eq. (\ref{Wmod}).
\item[(b)] Each copy $m\in [1,N_c]$ (in configuration $C_t[m]$ at time $t$) is cloned with rate $Y_{C_t[m]}$, i.e. if $\lfloor x \rfloor$ represents the integer part of $x$,
we generate a number $K_{C_t[m]}=\lfloor Y_{C_t[m]} \rfloor +1$ of identical clones with probability $Y_{C_t[m]} - \lfloor Y_{C_t[m]} \rfloor$, or
$K_{C_t[m]}=\lfloor Y_{C_t[m]} \rfloor$ otherwise. This cloning step includes the possibility of a copy leaving no offspring if $K_{C_t[m]}=0$.
\end{enumerate}
This procedure gives rise to a total of $N'_c(t)=\sum_{m=1}^{N_c} K_{C_t[m]}$ copies after cloning all of the original $N_c$ copies. To avoid numerical issues with the exponential explosion or implosion of the total population, we add a third, population-control step to the method described above:
\begin{enumerate}
\item[(c)] Once all copies have been evolved and replicated as dictated by their exit rates, the total number of copies $N'_c(t)$ is sent back to $N_c$ by an uniform cloning rate $X_t=N_c/N'_c(t)$.
\end{enumerate}
Fig.~\ref{sketchclon} sketches the cloning Monte Carlo algorithm. When the total number of clones $N_c$ \emph{is large enough} (and this is a key point in the algorithm), the global cloning factors $X_t$ are a good estimator of the average population increments in each time step, $X_t \approx \la {\cal N}(t-1)\ra/\la {\cal N}(t)\ra$, and hence we obtain an estimator of the dynamical free energy
\begin{equation}
\theta(\lambda) \approx -\frac{1}{t} \sum_{k=1}^t \ln X_k    \qquad \text{for } t\gg 1 \, ,
\label{musim}
\end{equation}
an expression which is expected to be exact in the limit of infinite number of clones, $N_c\to\infty$.

\subsection{Effects of a finite population of clones on the estimation of dynamical free energies}
\label{s2b}

Of course, in a real simulation we have at our disposal a hopefully large but still \emph{finite} number of copies $N_c$, and this limitation may drive the algorithm into troubles. A first problem is apparent: the cloning method will typically 
fail whenever the largest exit rate $Y_{C_t[m]}$ among the set of $N_c$ copies at a given time becomes of the order of $N_c$ itself \cite{hurtado09a}. If this is the case, configuration $C_t[m]$ will overpopulate all the other copies after the replication step (see Fig. \ref{sketchclon} and steps (b)-(c) in the algorithm of previous section), hence introducing a bias in the Monte Carlo sampling. 

The emergence of this problem depends on the magnitude of the current fluctuation, or equivalently on the value of conjugated parameter $\lambda$. The idea now is to estimate the critical $\lambda_c$ beyond which this main source of error becomes dominant, using tools from extreme value statistics \cite{sornette06a}. In order to to do so, we first define for a fixed parameter $\lambda$ the maximum exit rate among the set of $N_c$ copies at a given time $t$,
\be
Y_t^\text{max}\equiv \max (Y_{C_t[1]}, \ldots ,Y_{C_t[N_c]}) \, ,
\label{Ymax}
\ee
and consider the probability $P_{\lambda,<}^\text{max}(y)$ that  $Y_t^\text{max}$ is smaller than a threshold $y$. If $P_{\lambda}^>(y)$ is the probability that the exit rate of one clone is larger than $y$, and we assume statistical independence among the different clones, then $P_{\lambda,<}^\text{max}(y)$ can be simply written as
\be
P_{\lambda,<}^\text{max}(y)=[1-P_{\lambda}^>(y)]^{N_c} \, .
\label{PYmax}
\ee
In the limit of large $N_c$ the main contribution to $P_{\lambda,<}^\text{max}(y)$ is dominated by the tails of $P_{\lambda}^>(y)$, where this probability distribution is small, so in this large-$N_c$ limit we have
\be
P_{\lambda,<}^\text{max}(y)\simeq \exp[-N_c P_{\lambda}^>(y)] \, .
\label{PYmax2}
\ee
Consider now the following question: what is the value $y^*_\lambda(p)$ of the maximum which will not be exceeded with probability $p$? By definition, this value can be obtained from 
$p=P_{\lambda,<}^\text{max}[y^*_\lambda(p)]$, which according to Eq. (\ref{PYmax2}) leads to the following expression
\begin{equation}
P_{\lambda}^>[y^*_\lambda(p)]=\frac{1}{N_c} \ln\left(\frac{1}{p}\right) \, . \nonumber
\end{equation}
Now, for obvious reasons the maximum exit rate allowed by the algorithm is $N_c$ itself, and this limit leads to a critical value of the current-conjugated parameter $\lambda$ for a given confidence limit $p$ beyond which sampling problems in the cloning Monte Carlo algorithm will become evident \cite{hurtado09a}. To obtain this critical parameter $\lambda_c(p)$ (beyond which the maximum that will not be exceeded with probability $p$ is larger than $N_c$) we set $y^*_{\lambda_c}(p)=N_c$ in the previous identity, arriving at
\begin{equation}
P_{\lambda_c(p)}^>(N_c)=\frac{1}{N_c} \ln\left(\frac{1}{p}\right) \, .
\label{extremev2}
\end{equation}
This condition signals (with confidence level $p$) the onset of the systematic bias due to the finite number of clones in simulations, and defines the critical current-conjugated parameter $\lambda_c(p)$ beyond which this happens. To further proceed, we need the probability $P_{\lambda}^>(y)$ that the exit rate of a clone is larger than $y$, which implies some knowledge of the statistics at the end of a rare event \cite{hurtado09a}.

We hence consider the probability $P_t(C;Q)$ that the system is in configuration $C$ at time $t$ with a total time-integrated current $Q$. This probability obeys a simple recurrence relation (a sort of master equation) of the form
\begin{equation}
P_t(C;Q) = \sum_{C'} W_{C C'} P_{t-1}(C';Q-q_{C C'}) \, .
\label{mastereq}
\end{equation}
Iterating in time the previous relation, we arrive at
\begin{equation}
P_t(C;Q) = \sum_{C_{t-1}..C_0} \mathbb{W}_{C_0 \ldots C} \, \delta \left(Q - \sum_{k=0}^{t-1}q_{C_{k+1} C_k}\right) \, ,
\label{recurr}
\end{equation}
where we have used the path probability $\mathbb{W}_{C_0 \ldots C}= W_{C C_{t-1}}\ldots W_{C_1 C_0} P(C_0)$ for brevity, see also Eq. (\ref{ppath}). Note that $P_t(Q)=\sum_{C} P_t(C;Q)$ as expected, see Eq. (\ref{recurr1}) above. In this way, the probability measure of a configuration at the end of a  large deviation event of current $\hat{q}=Q/t$, or \emph{endtime statistics}, can be written as
\be
P_{\hat{q}}^{\text{end}}(C)= \frac{P_t(C;Q)}{P_t(Q)} \, .
\label{Pend}
\ee
Introducing now the moment-generating function of the distribution $P_t(C;Q)$,
\ben
Z_t(C;\lambda)&=&\sum_{Q} \text{e}^{\lambda Q} P_t(C;Q) \\
&=& \sum_{C_{t-1}\ldots C_0} \tilde{W}_{C C_{t-1}} \ldots \tilde{W}_{C_1 C_0} P(C_0)  \, , \nonumber
\label{piCend}
\een
such that $Z_t(\lambda)=\sum_C Z_t(C;\lambda)$, see Eq. (\ref{pi2}), one can show that for long times $t\gg 1$ \cite{hurtado14a,hurtado10a}
\be
P_{\lambda}^\text{end}(C) \equiv \frac{Z_t(C;\lambda)}{Z_t(\lambda)} = P_{{\hat{q}}_\lambda}^\text{end}(C)
\label{probCqend}
\ee
where ${\hat{q}}_\lambda=\partial_\lambda\theta(\lambda)$ is the conjugated current to parameter $\lambda$, which maximizes the Legendre transform (\ref{Legendre}). 
A direct inspection of Eq. (\ref{piCend}) at the light of the cloning Monte Carlo method of the previous section, see also Eq. (\ref{nclones}), shows that $P_{\lambda}^\text{end}(C)$ is proportional to the number of clones at a given time in configuration $C$ out of the total number of clones $N_c$.

Once $P_{\lambda}^\text{end}(C)$ has been defined, we can write the probability that a given clone has an exit rate $Y$ as
\be
P_\lambda(Y)=\sum_{C}  P_{\lambda}^\text{end}(C) \, \delta \left(Y - Y_C\right) \, ,
\ee
with $Y_C=\sum_{C'} \tilde{W}_{C' C}$. In this way
\begin{equation}
P_\lambda^>(y)=  \int_y^{\infty} dY \, P_\lambda(Y) = \sum_{C} P_{\lambda}^\text{end}(C) \, H \left(Y_C - y\right)\, , \nonumber
\end{equation}
where $H(x)$ is the Heaviside step function. Therefore the knowledge of endtime statistics during a large deviation event allows for an estimation of the range of validity of the cloning Monte Carlo method \cite{hurtado09a}. This calculation can be made completely explicit in particular models where analytical expressions for  the endtime distribution $P_{\lambda}^\text{end}(C)$ can be obtained \cite{hurtado09a}. In these cases it has been found that the probability distribution of exit rates exhibits power law tails, i.e. $P_\lambda(Y)\sim Y^{-\alpha(\lambda)}$ in the limit of large $Y$ which dominates the onset of the bias, with an exponent $\alpha(\lambda)$ which is typically a rational function of $\lambda$. This power law behavior is then inherited by the cumulative distribution $P_{\lambda}^>(y)\sim y^{-[\alpha(\lambda)-1]}$, and this can be used together with Eq (\ref{extremev2}) to arrive at an expression for the critical value of the current-conjugated parameter $\lambda_c(p)$ beyond which the bias due to the finite number of clones of the simulation appears, i.e.
\be
\alpha(\lambda_c)=2-\frac{\ln[\ln(p^{-1})]}{\ln(N_c)} \, .
\label{extremev3}
\ee
The accuracy of this expression, which bounds the range of validity of the cloning method, has been corroborated in numerical experiments of simple stochastic lattice gases \cite{hurtado09a}. Interestingly, the logarithmic dependence on the number of clones $N_c$ in the previous equation suggests that an appreciable increase of the range of validity of the algorithm in $\lambda$-space demands an exponential increase in the number of clones.

\subsection{Statistics and averages during a rare event}
\label{s2c}

In previous sections we have shown how to sample the dynamical free energy associated with the current (i.e. the Legendre transform of the current LDF) using a technique known as the cloning Monte Carlo method. Our aim in this paper is to apply this method to understand the physics of dynamical phase transitions, which appear as
singularities in the current LDF or the dFE and are accompanied by peculiar changes in the trajectories associated with rare events, which may include symmetry-breaking phenomena as captured by certain order parameter. These features can be only determined if we can measure observables of interest (as e.g. an order parameter, or a density profile) \emph{during} a large deviation, and this implies the characterization of the statistics of configurations in the course of a rare event. 

In order to define this statistics, we first introduce the probability $P_t(C_{t'},t';Q)$ that the system of interest was in configuration $C_{t'}$ at time $t'$ when at time $t$ the total integrated current is $Q$. Most importantly, time scales are such that $1 \ll t' \ll t$, so all times involved are large enough for the asymptotic large-deviation regime to be reached. To write down a formula for this probability, we just have to realize that this probability is nothing but the likelihood of all paths which traverse configuration $C_{t'}$ at time $t'$ such that the total accumulated current is $Q$. This reasoning leads to the following expression
\begin{widetext}
\begin{equation}
P_t(C_{t'},t';Q) = \sum_{C_{t}\ldots C_{t'+1} C_{t'-1} \ldots  C_0} W_{C_t C_{t-1}} \ldots W_{C_{t'+1} C_{t'}} 
W_{C_{t'} C_{t'-1}} \ldots W_{C_1 C_0} P(C_0) \, \delta \left(Q - \sum_{k=0}^{t-1} q_{C_{k+1} C_k}\right) \, .
\label{probCmid}
\end{equation}
Note that the sum over paths leaves fixed the configuration $C_{t'}$ at time $t'$. The probability of an arbitrary configuration $C$ during a large deviation event of the current ${\hat{q}}=Q/t$ (also known as \emph{midtime statistics}) can be now written as
\be
P_{\hat{q}}^\text{mid}(C) = \frac{P_t(C,t';Q)}{P_t(Q)} \, ,
\label{midtime}
\ee
and does not depend on the duration $t$ of the large deviation event and the time of measurement $t'$ as far as $1 \ll t' \ll t$. As before, if we define now the moment-generating function of the above distribution,
\be
Z_t(C_{t'},t';\lambda)=\sum_{Q} \text{e}^{\lambda Q} P_t(C_{t'},t';Q) = \sum_{C_{t}\ldots C_{t'+1} C_{t'-1} \ldots  C_0} \tilde{W}_{C_t C_{t-1}} \ldots \tilde{W}_{C_{t'+1} C_{t'}} \tilde{W}_{C_{t'} C_{t'-1}} \ldots \tilde{W}_{C_1 C_0} P(C_0)  \, ,
\label{piC}
\ee
it can be shown easily that the probability weight of configuration $C_{t'}$ at intermediate time $t'$ in a large deviation event of current ${\hat{q}}=Q/t$ can be written as \cite{hurtado14a,hurtado10a}
\be
P_{\hat{q}}^\text{mid}(C) =  \frac{P_t(C,t';Q)}{P_t(Q)} = \frac{Z_t(C,t';\lambda)}{Z_t(\lambda)} \equiv P_{\lambda}^\text{mid}(C)
\label{probCq}
\ee
for long times such that $1 \ll t' \ll t$, and where ${\hat{q}}$ and $\lambda$ are Legendre conjugated duals, ${\hat{q}}={\hat{q}}_\lambda=\partial_\lambda\theta(\lambda)$. To measure the typical value of an observable during a large deviation event, we have to define averages over the midtime statistics introduced above. In this way, if we denote our observable as $A(C)$, its average value during a large deviation event of the current with conjugated parameter $\lambda$ (i.e. with a current value of $\hat{q}_\lambda$) can be written as
\be
\la A\ra_\lambda^\text{mid} = \sum_C A(C)~ P_{\lambda}^\text{mid}(C) = \frac{1}{Z_t(\lambda)} \sum_{C_{t}\ldots C_0} W'_{C_t C_{t-1}} Y_{C_{t-1}} \ldots W'_{C_{t'+1} C_{t'}} Y_{C_{t'}} A(C_{t'}) W'_{C_{t'} C_{t'-1}} Y_{C_{t'-1}} \ldots W'_{C_1 C_0} Y_{C_0} P(C_0)  \, .
\label{piC}
\ee
\end{widetext}
This expression can be now conveniently read from a computational point of view to obtain an algorithm to measure averages of observables of interest during a large deviation event \cite{tailleur09a}. In particular, the idea is to run the cloning algorithm as described in previous sections, so that it generates  trajectories typical of a current fluctuation ${\hat{q}}_\lambda$. The difference now is that, when the time evolution reaches a chosen intermediate time $t'$, such that $1 \ll t' \ll t$, the value of the observable $A$ associated with the actual configuration is attached to each clone, see Eq. (\ref{piC}). In this way, whenever a clone is replicated, it carries a record of its value of observable $A$ at time $t'$. 
The midtime average of the observable of interest is now obtained from the average of the values of $A$ attached to each \emph{surviving} clone after a long time interval has passed since the measurement time $t'$ \cite{tailleur09a}. Since midtime statistics does not depend on $t$ or $t'$ as long as $1 \ll t' \ll t$, to better sample the midtime average it is convenient to nest in time several measurement-average pairs (characterized by a lag time $t_{\text{lag}}=t-t'$), as illustrated in Fig.~\ref{midtime}, leaving a reasonable waiting time $t_{\text{wait}}$ 
to avoid possible correlations between consecutive measurements.
Note that since usually $t_\text{wait}<t_\text{lag}$, several measurements must be carried out and transported \emph{in parallel} before averaging.
\begin{figure}[b]
\centerline{
\includegraphics[width=8.7cm]{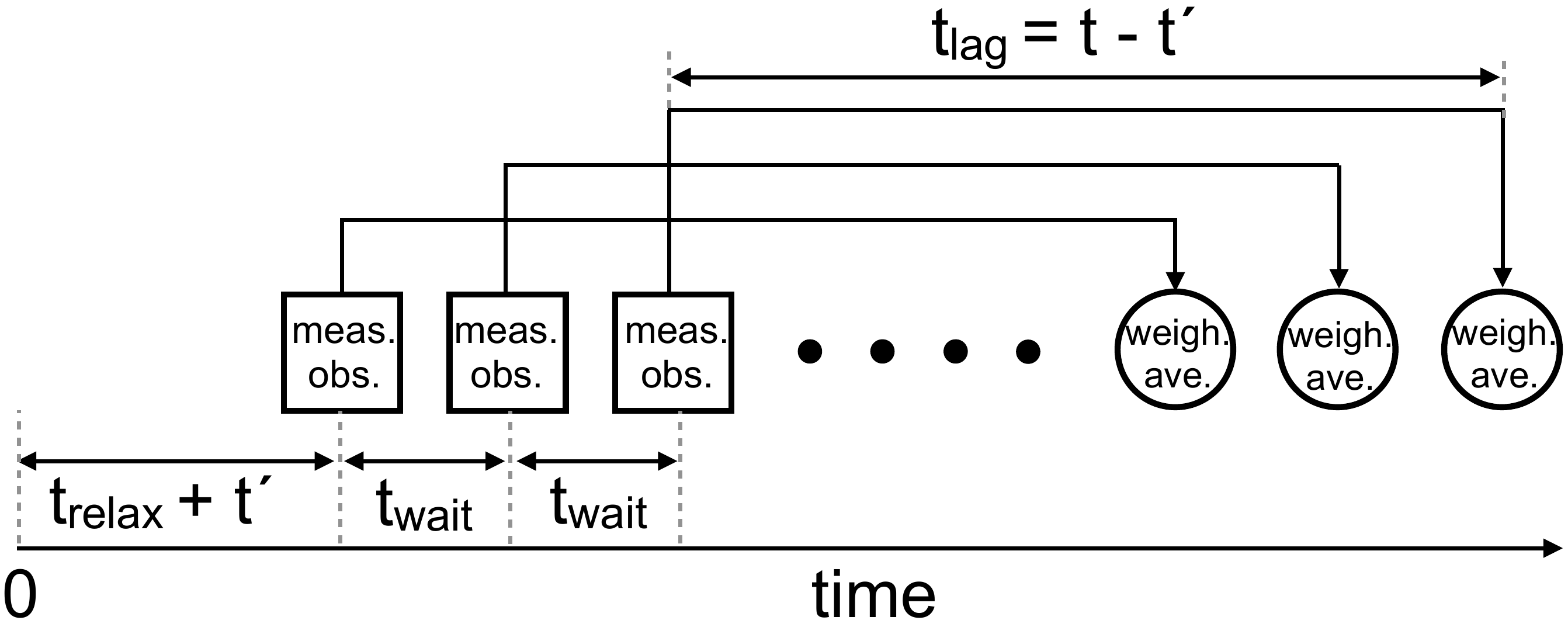}
}
\caption{Sketch of the midtime statistics of a rare event: At an intermediate time $t'$ the value of observable $A$ is measured and attached to each clone. The midtime average of $A$ is then taken after a long lag time and weighted according to the number of surviving clones attached to each measurement.}
\label{midtime}
\end{figure}

This midtime average, which includes the effect of the population cloning (and hence the exit rated $Y(C)$ along the different trajectories) is clearly different from the naive average of the observable $A(C)$ among the different clones at a given time, which yields the \emph{endtime statistics} associated with a large deviation event introduced in the previous section. 
It can be shown \cite{hurtado14a,hurtado10a,jack10a}, based on a spectral analysis of the deformed matrix with elements $\tilde{W}_{C' C}$, that endtime and midtime statistics during a large deviation event are different. Indeed midtime statistics captures the physics of the system during the rare event, while the endtime statistics only characterize transient behavior which appears both at the end and at the beginning of a rare event \cite{bertini15a,derrida07a,jack10a}.

\subsection{Continuous-time cloning algorithm}
\label{s2d}

\begin{figure}
\centerline{
\includegraphics[width=8cm]{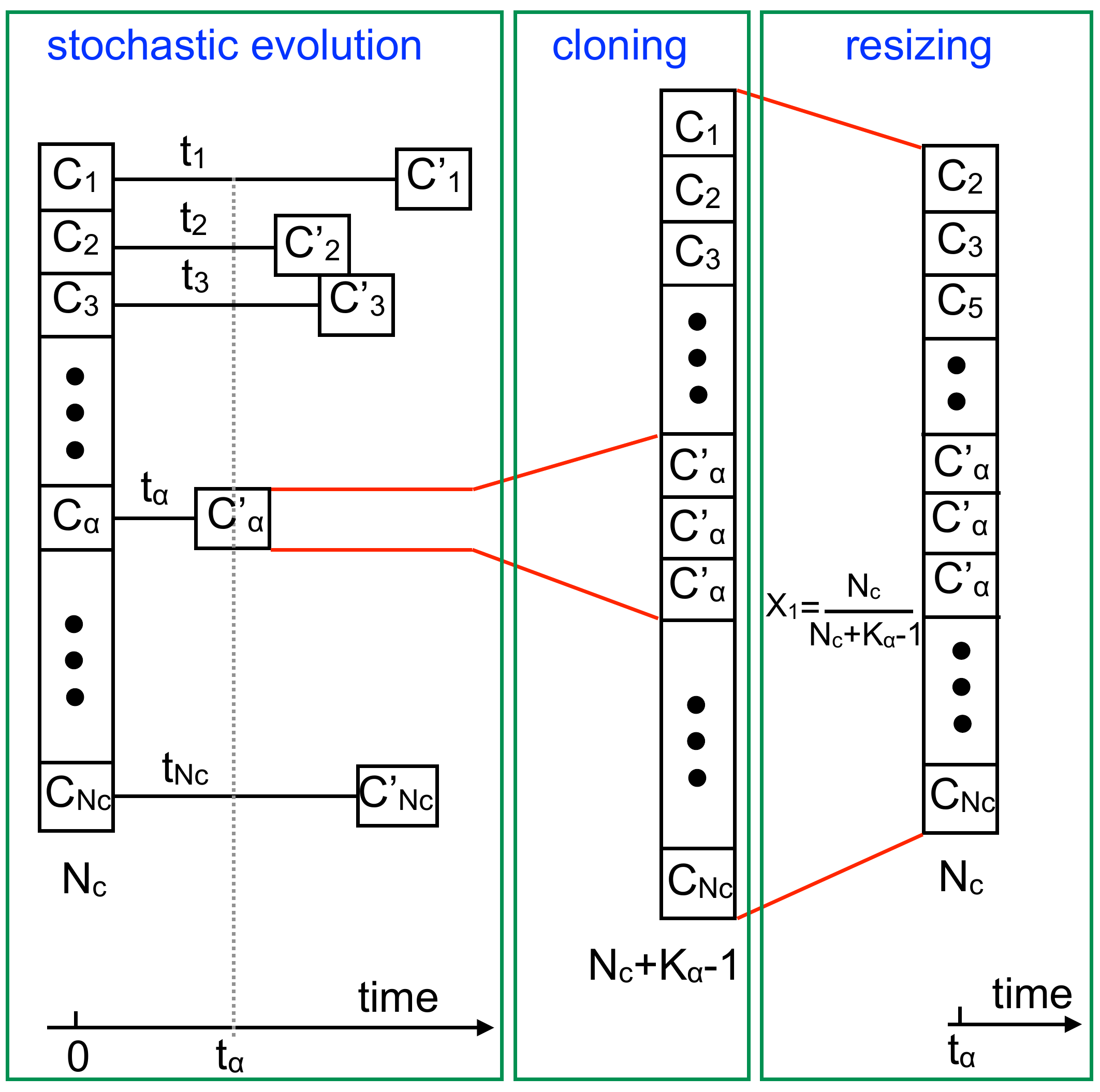}
}
\caption{Sketch of the cloning algorithm in continuous time during the evaluation of the large deviation function.}
\label{sketchclonCT}
\end{figure}

The previous time-discrete dynamics that allows for the measurement of time-integrated rare events can be readily extended to continuous-time dynamics. The evolution of the probability $P_t(C)$ of being in configuration $C$ at time $t$ is then governed by the master equation
$$
\partial_t P_t(C)=\sum_{C'\neq C}W_{CC'}P_t(C')-R(C)P_t(C)\, ,
$$
where $W_{CC'}$ is now the transition \emph{rate} from $C'$ to $C$ and $R(C)=\sum_{C'}W_{C'C}$ is the escape rate from configuration $C$. On the other hand, the probability of being in configuration $C$ having a time-integrated current $Q$ up to time $t$ is $P_t(C;Q)$, which evolves according to 
\beq
\partial_t P_t(C;Q)=\sum_{C'\neq C}W_{CC'}P_t(C',Q-q_{CC'})-R(C)P_t(C;Q)\, .
\label{evolPCQ}
\eeq
As in the discrete-time case, it is more convenient to work with the Laplace transform of $P_t(C;Q)$
$$
Z_t(C;\lambda)=\sum_Q \text{e}^{\lambda Q} P_t(C;Q) \, ,
$$
whose evolution is given, by virtue of Eq. \eqref{evolPCQ}, by 
\beq
\partial_t Z_t(C;\lambda)=\sum_{C'\neq C}\tilde{W}_{CC'} Z_t(C';\lambda)-R(C) Z_t(C;\lambda)\, .
\label{evolmodProb}
\eeq
Here $\tilde{W}_{CC'}=e^{\lambda q_{CC'}}W_{CC'}$ stands for modified transition rates, which can be much larger than the original $W_{CC'}$ for positive currents when $\lambda>0$ and for negative currents when $\lambda<0$. Summing over every configuration $C$ we get the dynamical partition function 
\beq
Z_t(\lambda)=\sum_C Z_t(C;\lambda)\asymp  \text{e}^{t\theta(\lambda)} 
\label{Zlambda}
\eeq
Thus, given $\lambda$ we could get $\theta(\lambda)$ by virtue of \eqref{Zlambda} just by simulating the evolution of $Z(C;\lambda)$ by means of \eqref{evolmodProb} up to time $t$ for many different realizations. However, this is not straightforward as Eq. \eqref{evolmodProb} is not a stochastic equation since $R(C)\neq \tilde{R}(C)\equiv \sum_{C'}\tilde{W}_{C'C}$. Nevertheless, it can be rewritten as a stochastic evolution which alternates with a cloning process
\beq
\begin{matrix} 
      \text{~~~~~~~~~~~~~~~~~stochastic evolution} \\ 
\partial_t Z_t(C;\lambda)=      \overbrace{ \sum_{C'\neq C}\tilde{W}_{CC'}Z_t(C';\lambda)-\tilde{R}(C)Z_t(C;\lambda)} \\
      +\underbrace{ \gamma(C) Z_t(C;\lambda)}\, , \\
      \text{cloning term}
   \end{matrix}   
   \label{contcloneq}
\eeq
with $\gamma(C)=\tilde{R}(C)-R(C)$. This dynamics can be computationally carried out by considering $N_c$ copies of the system which evolve as follows \cite{tailleur09a} (see Fig.~\ref{sketchclonCT}).\\ \\
\begin{enumerate}
\item[(a)] Set the time to $t_{\alpha}$, with $t_{\alpha}$ being the time of the first clone $\alpha$ to evolve.
\item[(b)] Change the configuration of copy $\alpha$ from configuration $C_{\alpha}$ to $C'_{\alpha}$ with probability $\tilde{W}_{C'_{\alpha}C_{\alpha}}/\tilde{R}(C_{\alpha})$. Compute the sojourn time $\Delta t$ of copy $\alpha$ in configuration $C'_{\alpha}$ until the next jump, distributed according to the Poisson law $P(\Delta t)=\tilde{R}(C'_{\alpha})e^{-\Delta t \tilde{R}(C'_{\alpha})}$.
\item[(c)] Clone configuration $C'_{\alpha}$ with rate $Y(C'_{\alpha})=e^{\Delta t \gamma(C'_{\alpha})}$, i.e.~generate a number $K_{\alpha}=\lfloor Y(C'_{\alpha})\rfloor + 1$ of identical clones with probability $Y(C'_{\alpha})-\lfloor Y(C'_{\alpha})\rfloor$; or $K_{\alpha}= \lfloor Y(C'_{\alpha})\rfloor$ otherwise. If $K_{\alpha}=0$, configuration $C'_{\alpha}$ is erased. 
\end{enumerate}
After a long time $t$, this cloning protocol changes the initial number of copies, resulting in an exponential growth or decrease of the population given by Eq. \eqref{Zlambda}. However, as we have to deal with a finite number of clones, we resize homogeneously the population to the original number of copies after each evolution:
\begin{enumerate}
\item[(d)] If $K_{\alpha}=0$ then one randomly chosen copy $\beta\neq \alpha$ is replicated, while if $K_{\alpha}>1$ then $K_{\alpha}-1$ copies, among the total $N_c+K_{\alpha}-1$, are randomly chosen and erased. 
\end{enumerate}
The resizing factor $X_1=N_c/(N_c+K_{\alpha}-1)$ is then stored at each time and the dynamical free energy can be computed as
$$
\theta(\lambda)\approx -\frac{1}{t}\sum_{k=1}^{M} \ln X_k\, ,
$$
where $M$ is the total number of jumps up to time $t$. Fig. \ref{sketchclonCT} sketches this procedure. \\

It is worth mentioning that one could alternatively analyse the above continuous-time system by simulating each clone for a fixed time $\Delta t$ and then using the discrete-time cloning method with $\Delta t$ as time step, see \cite{brewer18a} for details. However, the statistical errors strongly depend on $\Delta t$, but these can be considerably mitigated with a more efficient method of producing the new population by reducing the deletion of clones as described in \cite{brewer18a}.

\subsection{Recent improvements of the Monte Carlo cloning algorithm}
\label{s2e}

The cloning method described above in its discrete and continuous versions, despite being very successful, suffers from the fact that an exponential number of clones is needed to properly simulate trajectories associated with extremely rare events --corresponding to large values of $\lambda$--, as shown by \eqref{extremev3}, or in the vicinity of dynamical phase transitions. This has motivated recent improvements of the method aiming to mitigate such finite-population effects. One possibility recently explored consists in introducing an interpolation technique for the large deviation function based on the analysis of the systematic errors of a birth-death process \cite{nemoto17a}. An extension of such technique has led to a more efficient version of the cloning algorithm in continuous time which significantly improves the large deviation function estimators in the long time and large number of clones limit \cite{guevara17a}. In addition, some rigorous bounds on convergence properties of the algorithm have been established \cite{angeli18a,angeli19a}.

Another upgrade of the cloning method has been developed by incorporating to the population dynamics a controlling force which is determined by iterating a measurement-and-feedback scheme \cite{nemoto16a}. This method, recently extended to Markov jump processes \cite{nemoto17b}, is independent of the choice of the force in the limit of large number of clones. However, for a finite population of clones, it drastically improves the accuracy of the results, being necessary to simulate just one clone for the \emph{optimal} force, as in that case the cloning factor $\gamma(C)$ in Eq.\eqref{contcloneq} is constant, i.e.~independent of the configuration,  and equal to $\theta(\lambda)$. Thus, the dynamics associated with a rare fluctuation is stochastic and the rare trajectories of the original system correspond to the typical trajectories of the system under the action of the optimal force. The latter can be exactly obtained in terms of the eigenvector of the tilted generator associated with its largest eigenvalue $\theta(\lambda)$, but this is a very challenging task for many-particle systems, both numerically and analytically. Remarkably, the \emph{effective dynamics} under the optimal force corresponds to the Doob's h-transform \cite{jack10a,chetrite15a,chetrite15b}, which transforms the non-physical tilted dynamics into a proper stochastic dynamics. Nevertheless, despite the exact optimal force is difficult to obtain, combining an iterative estimation with the population dynamics has shown to be highly effective in finite-size scaling analysis of the first order dynamical phase transition present in the Fredrickson-Andersen model \cite{nemoto17b}.

A similar approach has been devised in Ref.~\cite{ray18a}, where the control strategy is performed by approximating the eigenvectors by means of guided distribution functions. In particular, the eigenvectors are approximated by product states of clusters of different number of sites. In this case, the iterative procedure to determine the eigenvectors is carried out outside the dynamics itself, in contrast to the approach of Refs.~\cite{nemoto16a,nemoto17b}. Analogous methods have been proposed in the context of the umbrella sampling technique \cite{oakes18a}, in which a reference dynamics is introduced to perform importance sampling of the rare trajectories of interest \cite{klymko18a}. In addition, besides the improvement in the copying and selection method pointed out at the end of the last section, further progress has been made regarding the parallel implementation of the cloning algorithm in \cite{brewer18a}, as well as its generalization to non-Markovian systems \cite{cavallaro16a}. Finally, it is worth mentioning that another recent algorithm using risk-sensitive and feedback control methods has been introduced to estimate the effective dynamics of a rare event from a single trajectory \cite{ferre18a}.

\section{Time-translational symmetry breaking at the trajectory level}
\label{s3}

Once equipped with the general computational tools described in previous sections to sample rare current fluctuations, we set out to study its application to explore different dynamical phase transitions which appear at the trajectory level. In particular, in this section we will use the cloning Monte Carlo method to characterize the DPT into a dynamical phase with broken time-translation invariance, appearing in two paradigmatic models of transport  for nonequilibrium physics, namely the Kipnis-Marchioro-Presutti (KMP) model of heat conduction \cite{kipnis82a} and the weakly asymmetric simple exclusion process (WASEP) \cite{derrida98a}.

\subsection{Ballistic energy packets in a model of heat conduction}
\label{s3a}

As we have already pointed out in the introduction, DPTs arise in the large size and/or long time limit and manifest as a non-analyticity in the LDF of time-averaged observables such as the current. Just as in standard critical phenomena, DPTs can be either discontinuous (first-order) or continuous (second-order).  The latter are associated with a symmetry breaking phenomenon, i.e.~the optimal path leading to a rare fluctuation breaks a symmetry of the corresponding large deviation functional \eqref{qLDF}. In this section we shall focus on an \emph{equilibrium} model of energy diffusion featuring a continuous DPT which breaks the time-translational invariance. 

In 1982 \cite{kipnis82a}, C.~Kipnis, C.~Marchioro and E.Presutti  introduced a simple stochastic model to understand in rigorous terms energy transport in systems with many degrees of freedom. This model, denoted hereafter as KMP model, has been paramount for the study of nonequilibrium phenomena since its formulation, leading to a number of new ideas and breakthroughs in this field. In particular, the authors of \cite{kipnis82a} were able to prove rigorously, starting from its microscopic Markovian dynamics, that the $1d$ KMP model in contact with two boundary thermal baths at different temperatures obeys Fourier's law. This is a key result in mathematical physics, as the microscopic foundations of Fourier's law still remain unknown in more realistic systems \cite{bonetto00a}.

\begin{figure}
\centerline{
\includegraphics[width=8.cm]{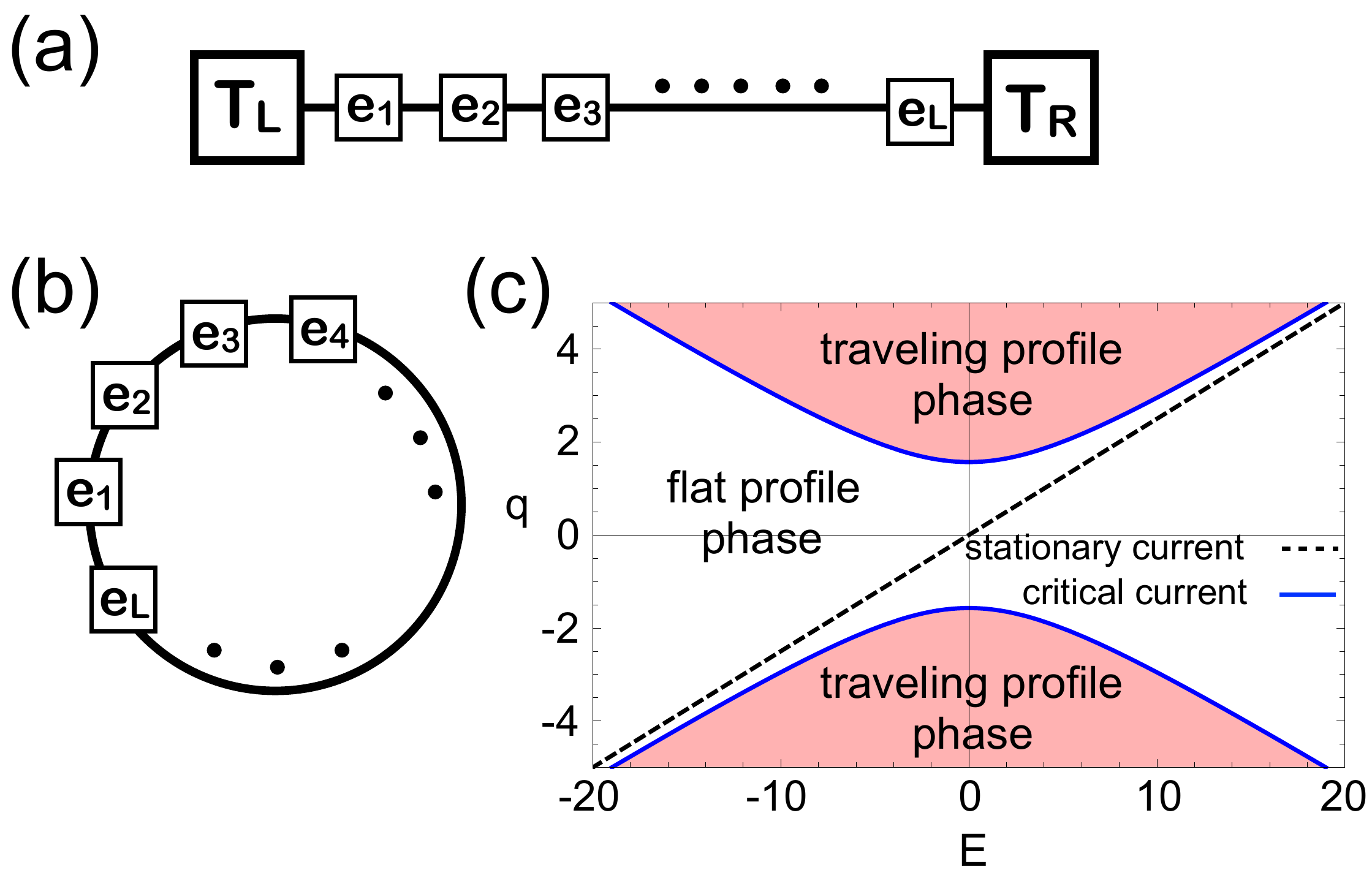}
}
\caption{(a) Sketch of the open $1d$ KMP model, where the extremal sites are attached to energy reservoirs at constant (but possibly different) temperatures. (b) Sketch of the periodic $1d$ KMP model. (c) Dynamical phase diagram of the periodic KMP model for $\rho_0=0.5$ as a function of a possible external field and the current.
}
\label{figmodelKMP}
\end{figure}

The KMP model consists of a $1d$ lattice of $L$ sites, 
see Fig. \ref{figmodelKMP}(a)-(b), though its definition can be generalized to different lattices in arbitrary dimension. The system's microscopic configuration is defined by a set $C\equiv \{e_i, i=1,\ldots,L\}$, where $e_i\in \mathbb{R}^+$ is the energy of site $i$. The dynamics is stochastic, and time can be discrete or continuous. In an elementary step, a pair of nearest neighbor sites $(i,i+1)$ is chosen at random, and the pair total energy is randomly redistributed locally so that the pair total energy is conserved in the interaction. In this way $(e_i,e_{i+1}) \to [e'_i(p),e'_{i+1}(p)]$ with
\ben
e'_i(p)  &=& p (e_i + e_{i+1}) \, , \label{KMP1} \\
e'_{i+1}(p)  &=& (1-p) (e_i + e_{i+1}) \label{KMP2} \, ,
\een
where $p$ is a uniform random number $p\in [0,1]$, so that $(e_i + e_{i+1})=[e'_i(p) + e'_{i+1}(p)]$ $\forall p$. 
With the aim of studying energy transport under a temperature gradient, KMP \cite{kipnis82a} coupled the model so defined to two boundary thermal baths at different temperatures, see Fig. \ref{figmodelKMP}(a). However, for our purposes here it is convenient to study a \emph{microcanonical} version of the KMP model, i.e. defined on a closed $1d$ lattice under periodic boundary conditions, see Fig. \ref{figmodelKMP}(b). In this case, the system is isolated and the total energy per particle, $\rho_0\equiv L^{-1} \sum_{i=1}^L e_i$, is conserved in the evolution. We note that in both cases the model can be generalized to include an external field $E$ driving the energy field is some preferred direction, though we will mostly focus on the $E=0$ KMP case below.

The macroscopic limit of this model is taken under diffusive scaling, with $x=i/L$ the macroscopic spatial variable and $t=m/L^2$ the macroscopic time ($m$ is the microscopic time variable), so the macroscopic energy density is defined as the average energy on each site; $\rho(x,t)=\la e_{i=Lx} \ra_{m=t L^2}$. 
In \cite{kipnis82a} it was shown by means of the duality technique \cite{carinci13a} that, at the macroscopic level, this model obeys Fourier's law with a diffusivity $D(\rho)=1/2$. As a consequence, when in contact with reservoirs at different temperatures $T_\text{L}\neq T_\text{R}$ at the left (L) and right (R) boundaries, the system reaches in the long time limit a stationary state characterised by a linear energy profile 
and a constant energy current, i.e. 
\ben
\rho_{\text{st}}(x) &=& T_{\text{L}} -x (T_{\text{L}}-T_{\text{R}}) \, , \\
q_{st} &=& -D(\rho_{st}) \partial_x \rho_{st}(x)=\frac{1}{2}(T_{\text{L}}-T_{\text{R}}) \, .
\een
On the other hand, when boundary conditions are periodic, i.e. in the absence of contact with thermal reservoirs, the KMP steady state is homogeneous, with a constant energy profile and no net current
\be
\rho_{\text{st}}(x) = \rho_0 \, , \qquad q_{st} = 0 \, .
\label{steady}
\ee
Moreover, the KMP mobility transport coefficient can be shown to be $\sigma(\rho)=\rho^2$ \cite{prados12a,hurtado14a}, and from these two transport coefficients, $D(\rho)$ and $\sigma(\rho)$, the macroscopic fluctuation theory (MFT) \cite{bertini15a} of Section \S\ref{s1} offers precise predictions on the fluctuating behavior of the KMP model, both under a temperature gradient \cite{hurtado09a,hurtado10a} and under periodic boundary conditions \cite{hurtado11a}.

For the case considered here, i.e. current statistics in the KMP model under periodic boundary conditions, MFT predicts a dynamical phase transition from (a) a phase for small current fluctuations around the average $q_{st} = 0$ where the Additivity Principle holds so the optimal profiles solution of the MFT problem are time-independent, $\rho_q(x,t)=\rho_0$ and $j_q(x,t)=q$ (see Section \S\ref{s1}), to (b) another dynamical phase for large current fluctuations where time-dependent optimal profiles emerge. Indeed, one can argue that small current fluctuations around the average result from the sum of local and uncorrelated jumps of the energy field, hence leading to Gaussian current statistics in agreement with the central limit theorem. However, for large enough currents beyond a critical threshold $|q_c|$, with
\be
|q_c|=\sqrt{\frac{8\pi^2 D^2(\rho_0) \sigma(\rho_0)}{\sigma''(\rho_0)}+\sigma^2(\rho_0)E^2}\, ,
\label{qcKMP}
\ee 
the system finds optimal to pack energy into a coherent traveling wave $\rho_q(x,t)=\omega_q(x-vt)$, moving at constant velocity $v$, which facilitates the rare event. Fig. \ref{figmodelKMP}(c) depicts the dynamical phase diagram for this second-order DPT, and includes the effect of a possible external field $E$ on the critical current \cite{hurtado14a}. For the particular case $E=0$ studied here, the appearance of ballistic energy packets for large current fluctuations is remarkable as it happens in an isolated (microcanonical) \emph{equilibrium} system, spontaneously breaking time-translational symmetry in $1d$. This is a beautiful example of the general observation that symmetry-breaking phase transitions forbidden in $1d$ equilibrium steady states can, however, emerge in the statistics of trajectories.

Our purpose in this section consists in characterizing the DPT in the KMP model using the cloning Monte Carlo method described in previous sections, providing some tips and tricks of the trade in the computation of large deviation functions, as well as details on order parameters for the transition and a characterization of the emergent structures. For simplicity we focus below on the discrete-time version of the cloning method as applied to the KMP model. If $C\equiv \{e_i, i=1,\ldots,L\}$ is the system configuration at a given time and 
\be
C_k^{(p)}\equiv \{e_1,\ldots,e'_k(p),e'_{k+1}(p),\ldots,e_L\} 
\ee
is the configuration resulting from $C$ after an interaction has occurred at pair $(k,k+1)$ with an exchange parameter $p$, see Eqs. (\ref{KMP1})-(\ref{KMP2}), then the transition probability between these two configurations is simply $W_{C_k^{(p)},C}=L^{-1}$, as the interaction probability of all $L$ pairs is the same and $p$ is an homogeneous random number in the unit interval. We now define the energy current involved in this transition as the energy traveling to the right divided by the total number of pairs, namely
\be
q_{C_k^{(p)},C}= \frac{1}{L}[e_k - e'_k(p)] = \frac{1}{L}[e_k - p (e_k + e_{k+1})] \, ,
\label{currKMP}
\ee
so the modified transition probability $\tilde{W}_{C_k^{(p)},C}$ defined in the cloning Monte Carlo method, see Eq. (\ref{moddyn}), can be written as
\be
\tilde{W}_{C_k^{(p)},C} = \frac{1}{L} \text{e}^{\frac{\lambda}{L}[e_k - p (e_k + e_{k+1})]} \, .
\label{moddynKMP}
\ee
The associated exit rate $Y_C$, see Eq. (\ref{Yc}), hence reads
\ben
Y_C &=& \sum_{C_k^{(p)}} \tilde{W}_{C_k^{(p)},C} = \sum_{k=1}^L \int_0^1 dp ~\tilde{W}_{C_k^{(p)},C} \nonumber \\
&=&  \sum_{k=1}^L \frac{\text{e}^{\frac{\lambda}{L}e_k} - \text{e}^{-\frac{\lambda}{L}e_{k+1}}}{\lambda(e_k + e_{k+1})} \, ,
\label{YcKMP}
\een
so the normalized modified dynamics to study current statistics in the $1d$ periodic KMP model is just 
\be
W'_{C_k^{(p)},C} = \frac{1}{L Y_C} \text{e}^{\frac{\lambda}{L}[e_k - p (e_k + e_{k+1})]} \, ,
\ee
see Section \S\ref{s2a}. Note that this modified dynamics weights each possible transition according to the current involved in the step and the intensive parameter $\lambda$ conjugated to the current. This new stochastic dynamics can now be used in conjunction with the clone population replication/pruning scheme of \S\ref{s2a} to sample the current large-deviations statistics in this model.

\begin{figure}
\centerline{
\includegraphics[width=8.cm]{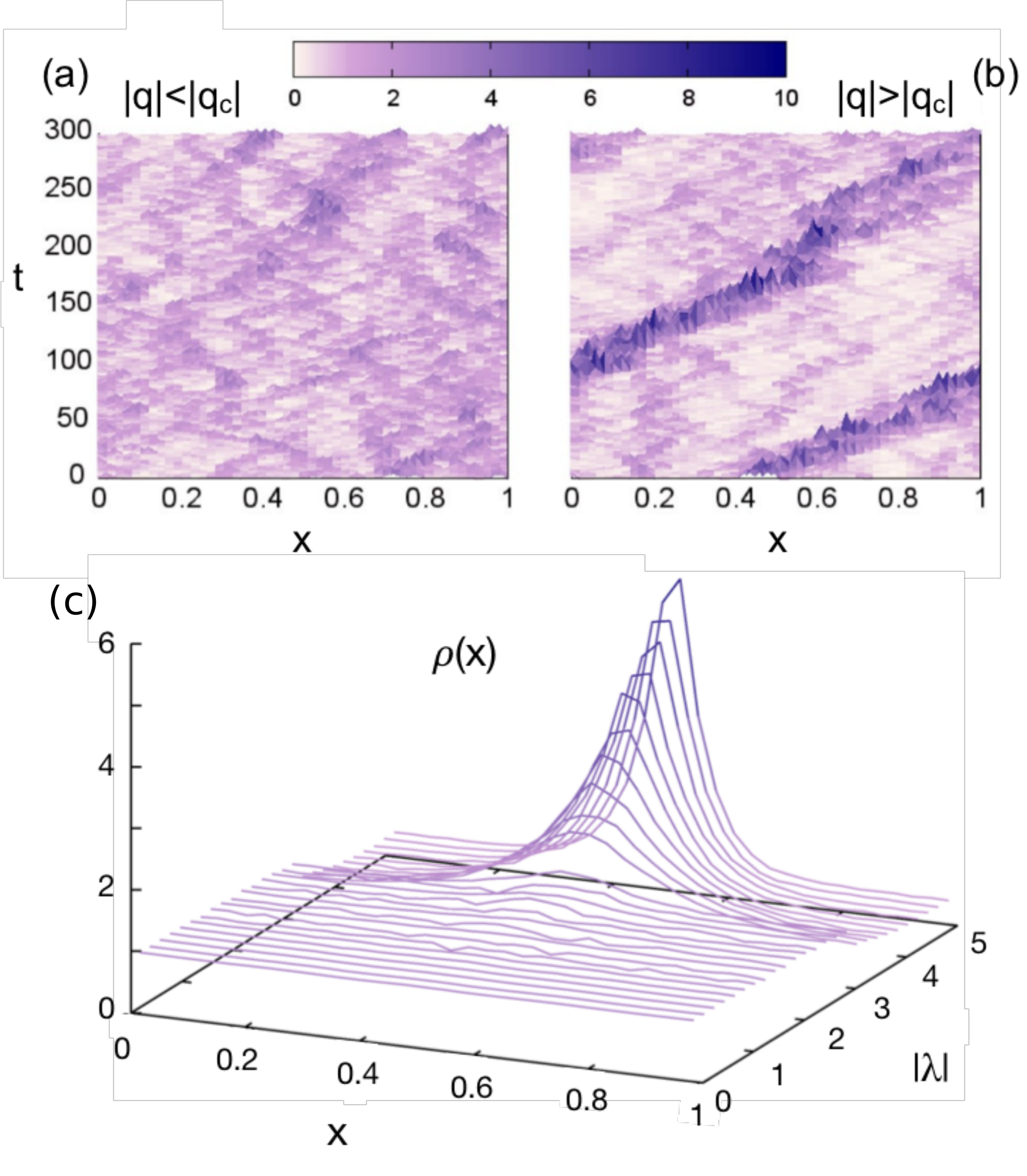}
}
\caption{Results for the periodic KMP model with $E=0$ and $\rho_0=1$: (a) Trajectories associated with subcritical current fluctuations $|q|<|q_c|$, where no spatial structure is observed. (b) Trajectories corresponding to currents beyond the critical $|q|>|q_c|$, where energy is ballistically transported. 
(c) Optimal energy profiles, obtained with midtime statistics, for different current fluctuations and $L=32$.
}
\label{figmodelKMPtraj}
\end{figure} 

A first evidence of the DPT phenomenon described above for the KMP model is expected to appear at the level of individual trajectories of the energy field for current fluctuations below and beyond the critical current $|q_c|$. With this aim we simulated the KMP model on a $1d$ ring lattice using the cloning method for two different values of the conjugated parameter $\lambda$, such that in one case the current $q=q_\lambda$ associated with this $\lambda$ was below the critical threshold, $q<|q_c|$, while in the other case $q>|q_c|$. Figs. \ref{figmodelKMPtraj}(a)-(b) show the resulting typical trajectories in both cases as a spatiotemporal raster plot of the energy field. 
In particular, Fig. \ref{figmodelKMPtraj}(a) shows that the system can sustain a net current fluctuation $q<|q_c|$ in the absence of any macroscopic structure, i.e. with an homogeneous average density profile. However, Fig. \ref{figmodelKMPtraj}(b) clearly demonstrates that in order to sustain 
a large energy current ($|q|>|q_c|$) the system accumulates energy in the form of packets that propagate ballistically (with a constant velocity on average) across the lattice. By contrast, trajectories producing moderate currents do not feature any spatial structure. 

A more quantitative characterization of these energy packets for $|q|>|q_c|$ can be obtained by measuring their average shape. Due to the packet's ballistic motion and the lattice periodicity, a naive averaging of density profiles along multiple rare-event trajectories would miss the packet structure, leading instead to homogeneous density profiles. In order to correctly measure the traveling-wave energy profile and not to blur away its spatial structure, we performed averages of the density field around its instantaneous \emph{center of mass}. 
Indeed, due to the periodicity of the $1d$ lattice we can consider the system as a ring embedded in a two-dimensional space, and this allows us to assign an angle $\theta_i=2\pi i/L$ to each site $i\in[1,L]$ in the lattice. In this way we define now the angular position of the energy field center of mass for a given configuration $C\equiv \{e_i, i=1,\ldots,L\}$ as $\theta_\text{CM}= \tan^{-1}(y_\text{CM}/x_\text{CM})$,
where 
\ben
x_\text{CM} &=& \frac{1}{L\rho_0} \sum_{i=1}^L e_i \cos \theta_i \, , \label{xcm}\\
y_\text{CM} &=& \frac{1}{L\rho_0} \sum_{i=1}^L e_i \sin \theta_i \, . \label{ycm}
\een
Hence, every time we have to average configurations during a large deviation current event, we rotate the system by an angle $\theta_\text{CM}$ before averaging so that the angular center of mass always sits in the origin. 
In this way we correctly capture the spatial structure of the traveling-wave profile, which would otherwise fade away. Some caution is needed however: this measuring method leads to spurious weakly-structured energy profiles for subcritical current fluctuations $|q|<|q_c|$, as averaging 
random homogeneous energy profiles around their random center of mass results in a weak but non-trivial spatial structure.
This potential problem can be easily fixed by noting that this spurious profile is of course independent of the current $q$ for $|q|<|q_c|$, so it can be easily subtracted. 
Fig.~\ref{figmodelKMPtraj}(c) shows the measured average optimal energy profile responsible for a current fluctuation $q_\lambda$ obtained with this technique as a function of the conjugate parameter $|\lambda|$. The appearance of a critical value $\lambda=\lambda_c$, or equivalently a critical current $q_c$, where the optimal profile changes from homogeneous to structured clearly signals the onset of the DPT into the traveling-wave dynamical phase. These measurements were performed for a system with $L=32$ sites using a large number of clones $N_c=10^4$. 

\begin{figure}
\centerline{
\includegraphics[width=8.cm]{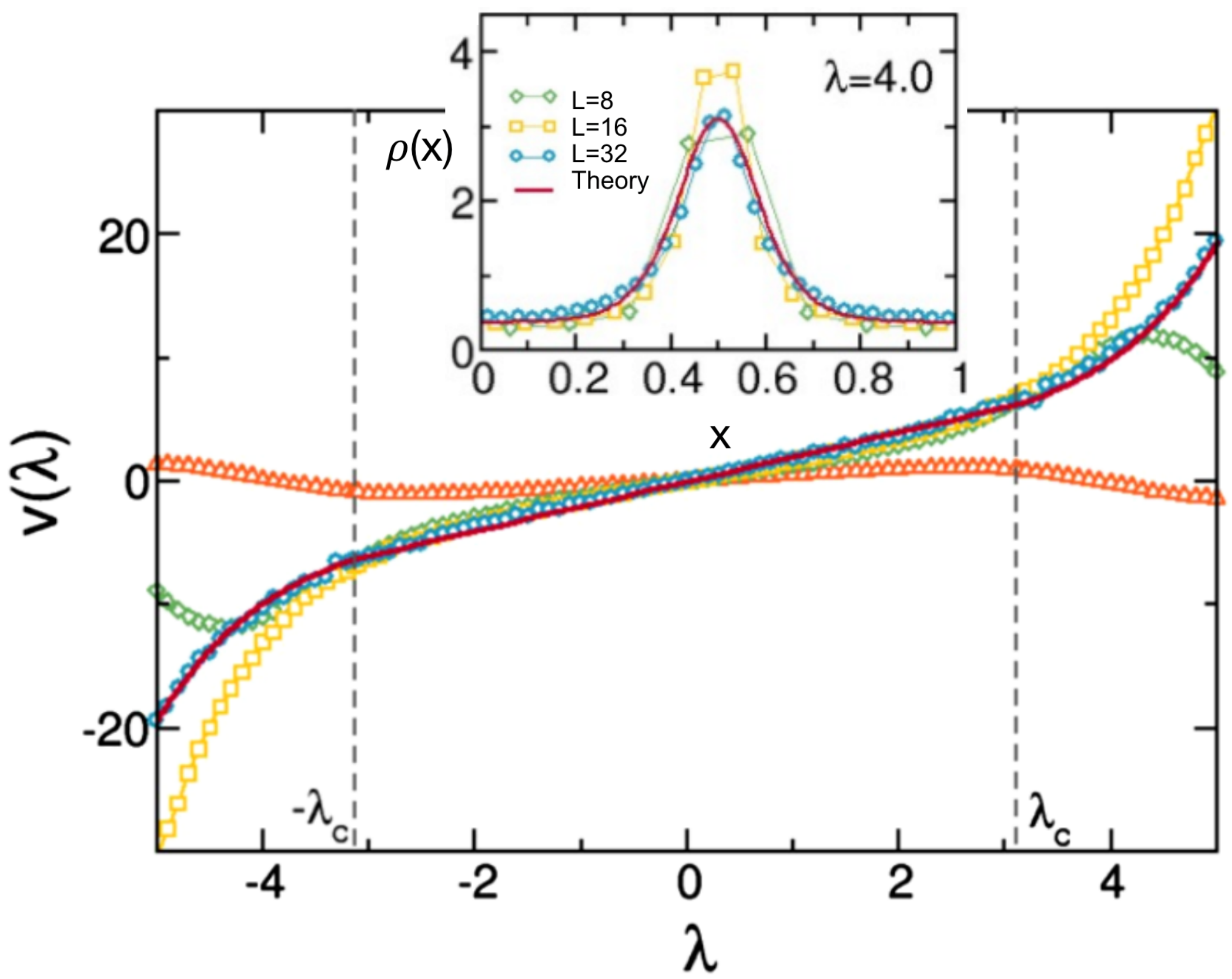}
}
\caption{Velocity of the center of mass motion measured as a function of the current-conjugate parameter $\lambda$ for increasing values of $L$, and MFT theoretical prediction. In these simulation the number of clones is $N_c=10^4$.
Inset: Finite-size corrections affecting the shape of the traveling profile for $\lambda=4$. The red solid line corresponds to the theoretical prediction while symbols are simulation results for different system sizes ($L=8, 16, 32$). 
}
\label{figmodelKMPtraj2}
\end{figure}

As described in Section \S\ref{s2b}, these measurements are strongly affected by finite size effects both on the system sizes and the number of clones employed in the sampling, so a careful analysis of how $L$ and $N_c$ affect the measurements is always needed. Indeed, in the inset to Fig.~\ref{figmodelKMPtraj2} we display the measured energy packet profile for a current fluctuation in the traveling-wave phase ($\lambda=4>\lambda_c$) for increasing system sizes $L$. We observe how the measured profiles suffer from the strong finite-size effects anticipated above, but for $L=32$ there is already a very good agreement with the optimal energy profiles predicted by MFT \cite{hurtado11a,hurtado14a}. 
A similar analysis can be performed as a function of the number of clones $N_c$, and this suggests that $N_c=10^4$ suffices in this case to sample accurately current statistics and the associated structures across the DPT. As we will see below, these finite-population effects increase in higher dimensions, demanding the simulation of very large numbers of clones in parallel. 

Another key feature of the supercritical energy packets is their propagation velocity. To measure this observable, we define a \emph{quasi-instantaneous} velocity for a given trajectory as the average slope of the packet peak trajectory in a space-time diagram for a short time interval, see Fig. \ref{figmodelKMPtraj}(b). In particular, by fitting the motion of the center of mass during small time intervals $\Delta t$ to a ballistic law, $r(t+\Delta t) - r(t) = vt$, see Fig. \ref{figmodelKMPtraj}(b), and using this observable to obtain midtime averages using the sampling scheme of Section \S\ref{s2c}, we obtained the data plotted in the main panel of Fig. \ref{figmodelKMPtraj2}. In particular, we plot in this graph the velocity of the center of mass motion as a function of the bias parameter $\lambda$ for different system sizes, as well as the macroscopic theoretical prediction based on MFT calculations \cite{hurtado11a}. An excellent agreement is found already for $L=32$, while clear finite-size corrections are observed below this size, as in other observables above. Note that while the center of mass velocity is linear in $\lambda$ in the subcritical region (i.e. this velocity is just proportional to the current), for currents beyond the critical threshold, or equivalently $|\lambda|>|\lambda_c|$, the velocity $v(\lambda)$ becomes a nonlinear monotonously increasing function of $\lambda$, another trait of the DPT.

\begin{figure}
\centerline{
\includegraphics[width=8.cm]{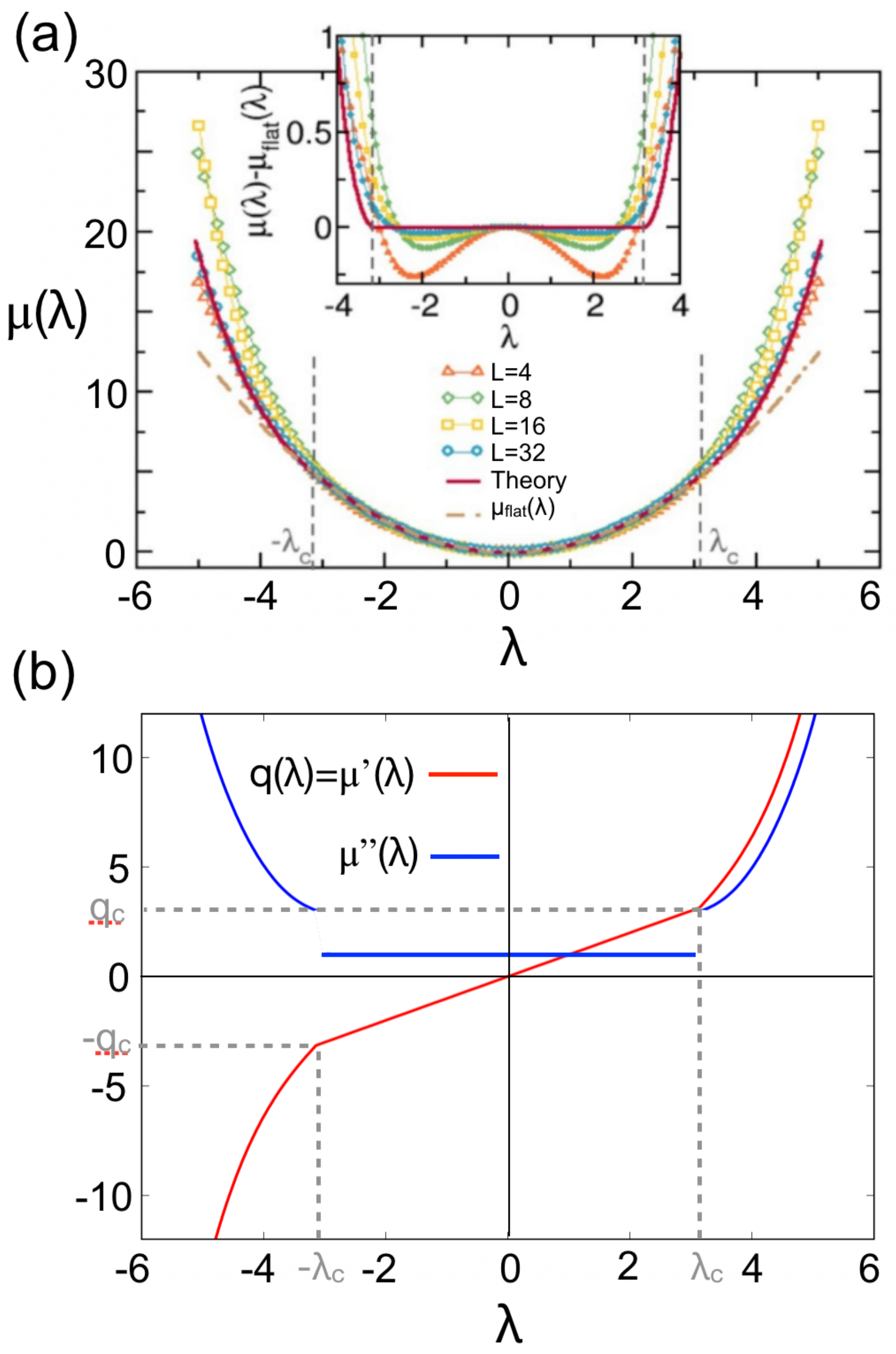}
}
\caption{(a) Dynamical free energy $\mu(\lambda)$ for the periodic KMP model with $E=0$ and $e_{tot}=1$. Red solid line corresponds to the theoretical predictions, while symbols are numerical results for different sizes ($L=4, 8, 16, 32$). Inset: Difference between $\mu(\lambda)$ and the prediction for Gaussian fluctuations $\mu_{\text{flat}}(\lambda)$   (b) First and second derivatives of the dynamical free energy.
}
\label{figdatKMP}
\end{figure} 

Finally, the cloning algorithm permits to measure not only the typical profiles giving rise to a large dynamical fluctuation but also its dynamical free energy $\mu(\lambda)$, which is nothing but the Legendre transform of the current LDF $G(q)$, with $\lambda$ being the parameter conjugate to the current $q$. In Fig.\ref{figdatKMP}(a) we present the theoretical $\mu(\lambda)$ along with the numerical values measured for different system sizes and $N_c=10^4$. We clearly observe how for currents below the critical one ($|\lambda|<|\lambda_c|$) the Gaussian solution --associated with the time-independent homogeneous profiles-- is the minimizing one, while for currents above the critical threshold ($|\lambda|>|\lambda_c|$) the dFE branch associated with traveling-wave profiles is the optimal one [see inset to Fig.~\ref{figdatKMP}(a)]. Note that in virtue of the Legendre transform connecting $G(q)$ and $\mu(\lambda)$ given by \eqref{LegendreMacro}, we get that $P_{\rm MFT}(q)>P_{\rm flat}(q)$ for $|q|>|q_c|$ implies that $\mu_{\rm MFT}(\lambda)>\mu_{\rm flat}(\lambda)$ for $|\lambda|>|\lambda_c|$, as observed in Fig.~\ref{figdatKMP}(a). This highlights the fact that large current fluctuations are far more probable --via the emergence of complex spatio-temporal structures-- than what is predicted naively by Gaussian statistics. The transition between both regimes corresponds to a second-order DPT as $\mu(\lambda)$ presents a discontinuity in its second derivative, i.e. $\lim_{|\lambda| \to |\lambda_c|^-}\partial^2_{\lambda}\mu(\lambda)\neq \lim_{|\lambda| \to |\lambda_c|^+} \partial^2_{\lambda}\mu(\lambda)$. This is displayed in Fig.~\ref{figdatKMP}(b) as obtained from MFT \cite{hurtado11a}, where we also present its first derivative $\mu'(\lambda)$ corresponding to the current $q_\lambda$ associated with each $\lambda$.

\begin{figure}
\centerline{
\includegraphics[width=8cm]{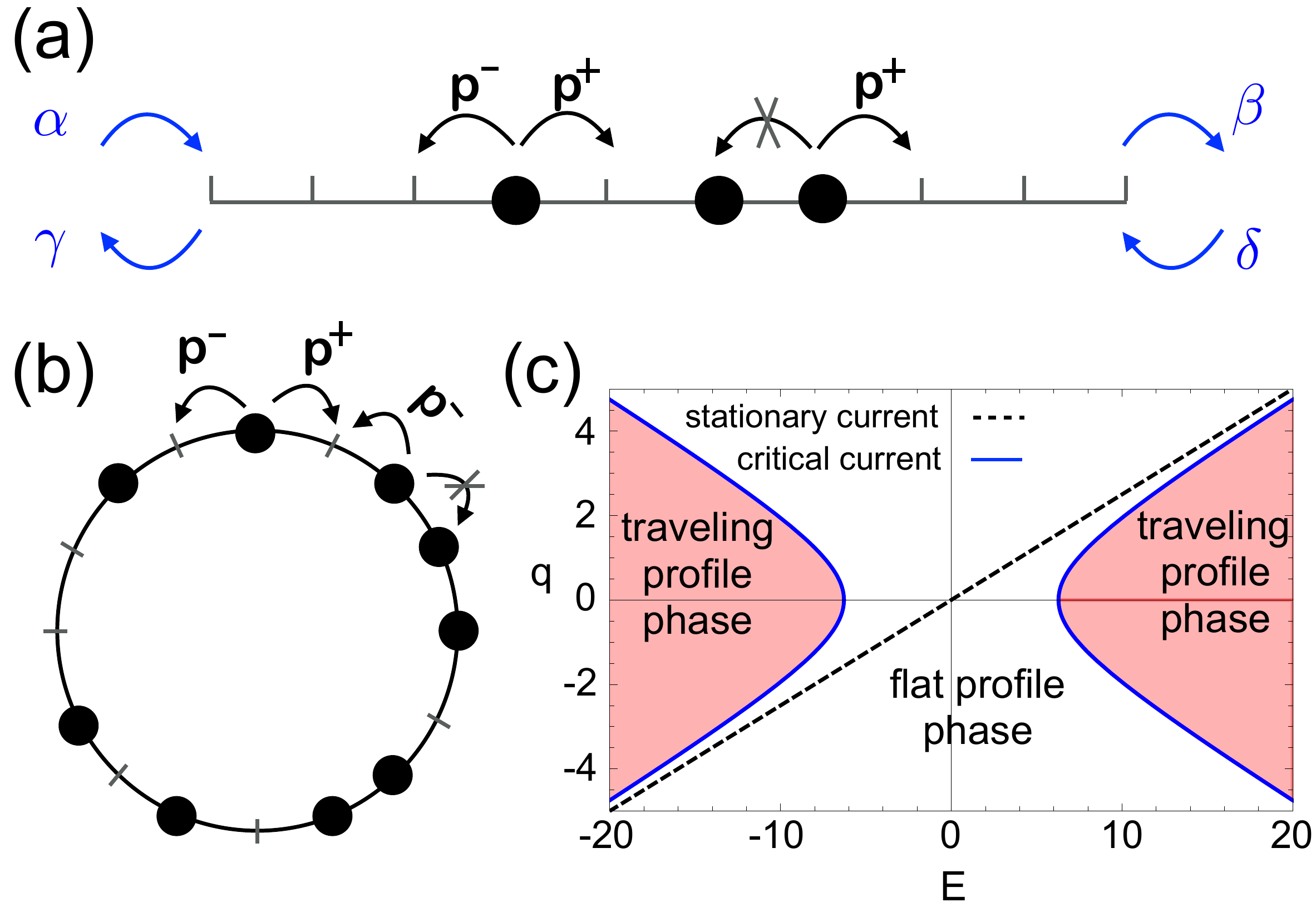}
}
\caption{(a) Sketch of the open $1d$ WASEP, where particles are transferred to the system through its boundaries. (b) Sketch of the periodic $1d$ WASEP. (c) Dynamical phase diagram of the periodic WASEP for $\rho_{0}=0.5$ as a function of the external field and the current.
}
\label{figmodelSSEP}
\end{figure}

\subsection{Jammed trajectories in a model of particle diffusion}
\label{s3b}

Another archetypical stochastic lattice model which displays an interesting DPT is the weakly asymmetric simple exclusion process (WASEP) \cite{gartner87a,demasi89a}, a schematic model of particle diffusion. This model exhibits a DPT in its current statistics which is mathematically very similar to that of the KMP model described above \cite{bodineau05a,perez-espigares13a}. However, as we will see below, the DPT in WASEP has a markedly different physical interpretation, which again gives valuable insights into the range of possibilities that an apparently simple system has at its disposal in order to maximize the probability of
a given fluctuation. Moreover, understanding this DPT in $1d$ will help us to better understand the important effect of dimension on the current statistics of driven system, which will be the focus of the Section \S\ref{s4} below.

The WASEP is defined on a $1d$-lattice of $L$ sites, each of which can be either empty or occupied at most by one particle, see Figs. \ref{figmodelSSEP}(a)-(b). As for the KMP model, the WASEP can be defined in arbitrary dimensions and different types of lattices (indeed we will study a $2d$ case in next section). A microscopic configuration is defined as $C \equiv  \{ n_i, i=1,\ldots L\}$, where $n_i=0,1$ is the occupation number of site $i$ at a given time, which as before can be either discrete or continuous. Every particle can randomly jump to one of its nearest neighboring sites, provided this is empty. The jump rate is $p^{\pm}=pe^{\pm E/L}$, for jumps to the right ($+$) or left ($-$) neighbor, with $p$ an overall jump rate which sets the microscopic time scale, and $E$ an external field biasing the motion along a given direction. The term \emph{weakly} in WASEP denotes the fact that the rate asymmetry, $p^+-p^-$, scales as $L^{-1}$ for large $L$, allowing the external field to compete on equal footing with the particle diffusion process.
As in the KMP model of previous section, we may choose to (a) couple the WASEP model to two particle reservoirs at the boundaries, which may work at different chemical potentials (or densities) thus inducing an external gradient, or rather (b) impose periodic boundary conditions so the total number of particles in the system $M\equiv \rho_0 L=\sum_{i=1}^L n_i$ is conserved during the evolution. To implement boundary reservoirs, 
particles are injected to the first (last) site at rate $\alpha$ ($\delta$) provided it is empty and removed from the first (last) site at rate $\gamma$ ($\beta$) provided it is occupied [see Fig.~\ref{figmodelSSEP}(a)]. This mechanism mimics the effect of boundary
reservoirs at densities $\rho_\text{L}=\alpha/(\alpha + \gamma)$ and $\rho_\text{R}=\delta/(\delta + \beta)$. On the other hand, removing the extremal reservoirs and introducing periodic boundary conditions we obtain the microcanonical version of the WASEP that will be the focus of study in this section. The stationary state under periodic boundaries has a homogeneous density profile $\rho_{st}(x)=\rho_0=M/L$ equal to the average density in the system, and a net current $q_{st}(E)=\sigma(\rho_0) E = 2 p \rho_0(1-\rho_0) E$ due to the presence of the weak external field.

\begin{figure}
\centerline{\includegraphics[width=9cm]{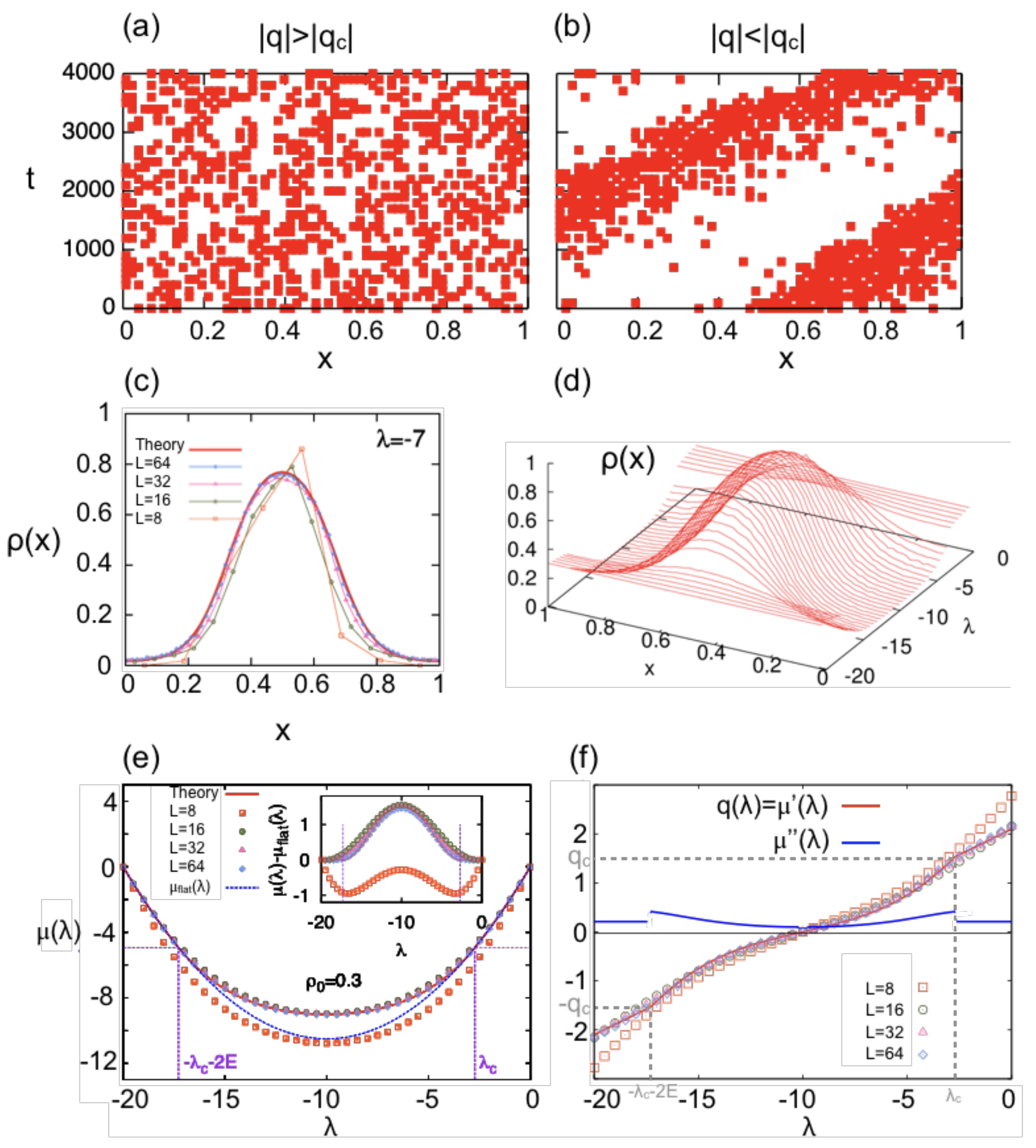}}
\caption{Results for the periodic WASEP with $E=10>E_c$ and $\rho_0=0.3$: (a) Trajectory associated with a supercritical current fluctuations $|q|>|q_c|$, where no spatial structure is observed. (b) Trajectory corresponding to a current beyond the critical one, $|q|>|q_c|$, where particles self-organize into a traveling jammed state. (c) Average shape of the traveling-wave profile for $\lambda=-7$. The red solid line corresponds to the theoretical prediction while symbols are simulation results with midtime statistics for different system sizes ($L=8, 16, 32, 64$). (d) Measured optimal energy profiles with midtime statistics for different current fluctuations (as parametrized by $\lambda$) and $L=64$. (e) Measured dynamical free energy $\mu(\lambda)$.
The red solid line corresponds to the theoretical prediction, while symbols are numerical results for different sizes ($L=8, 16, 32, 64$). Inset: Difference between $\mu(\lambda)$ and the quadratic approximation for Gaussian fluctuations. 
(f) First and second derivatives of the dynamical free energy, and comparison with data. In all cases the cloning simulations used a total of $N_c=2\times 10^4$ clones to sample large deviation statistics.
}
\label{fig_ssb_WASEP}
\end{figure}

Averaging over trajectories on each site and applying the diffusive scaling limit, we can define a macroscopic density field for WASEP as $\rho (x,t)=\la n_{i=Lx} \ra_{m=tL^2}$, with $m$ the microscopic time. In this mesoscopic limit, the system evolution is captured by a fluctuating hydrodynamics equation, see Eq. (\ref{current}) in Section \S\ref{s1}, defined in terms of two transport coefficients, the diffusivity $D(\rho)$ and the mobility $\sigma(\rho)$, which for WASEP are $D(\rho)=p$ and $\sigma(\rho)=2p\rho(1-\rho)$ \cite{spohn12a}.
Using these two transport coefficients, the macroscopic fluctuation theory of Section \S\ref{s1} predicts a DPT in the current statistics of the microcanonical $1d$ WASEP to a dynamical phase dominated by traveling-wave density profiles, with a mathematical structure very similar to that of the DPT in the KMP model but with several key differences. 
For instance, the critical current where the DPT appears has the same general form as in the KMP model, see Eq. (\ref{qcKMP}), namely
\be
|q_c|=\sqrt{\frac{8\pi^2 D^2(\rho_0) \sigma(\rho_0)}{\sigma''(\rho_0)}+\sigma^2(\rho_0)E^2}\, .
\label{qcWASEP}
\ee 
However, and most importantly, while $\sigma''(\rho_0)>0$ for the KMP model, for the WASEP we have $\sigma(\rho)=2p\rho(1-\rho)$ so that $\sigma''(\rho)=-4p<0$. This key difference, which originates in the particle exclusion --a main feature of the WASEP--, immediately introduces a critical value for the external field, 
\be
|E_c|=\pi/\sqrt{2p\rho_0(1-\rho_0)} \, ,
\label{EcWASEP}
\ee
below which no DPT is possible as the critical current $|q_c|$ becomes complex. Fig. \ref{figmodelSSEP}(c) shows the resulting dynamical phase diagram for current statistics in $1d$ WASEP as derived from MFT, see Eq. (\ref{qcWASEP}). In this way we can write
\be
|q_c|=|q_{st}(E)|\sqrt{1-\left(\frac{E_c}{E}\right)^2}\, ,
\label{qcWASEP2}
\ee
so clearly $|q_c|<|q_{st}|$ for WASEP under external fields of magnitude larger than the critical one. This is another important difference between the KMP and WASEP DPTs.
In particular, for the KMP model there is no current in the steady state (in the absence of external field) and the DPT appears for \emph{large enough currents}, where the system facilitates the fluctuation by localizing energy in packets that travel ballistically across the system. On the other hand, and in stark contrast, the DPT in WASEP appears for current \emph{below} the average stationary current. In this case the dynamic phase transition corresponds to the emergence of a \emph{macroscopic jammed state} which hinders transport of particles to facilitate a current fluctuation well below the average.

To characterize this sort of jamming transition at the trajectory level, we performed Monte Carlo cloning simulations to unveil the current statistics of the $1d$ periodic WASEP \cite{perez-espigares13a} for an average density $\rho_0=0.3$ under a supercritical external field $E=10>E_c$, and for increasing system sizes $L\in [8, 64]$. Fig. \ref{fig_ssb_WASEP} summarizes our results, that we obtained using $N_c=2\times 10^4$ clones evolving in parallel. In particular, Figs. \ref{fig_ssb_WASEP}(a)-(b) depict spatiotemporal raster plots of trajectories obtained in cloning simulations for currents larger and smaller than the critical current $|q_c|$. The trajectory associated with a current fluctuation $|q|>|q_c|$ is homogeneous and apparently random, with no clear structure, while for $|q|<|q_c|$ particles are localized in a jammed region which moves at constant velocity across the system. The average shape of the jammed density profile was measured in detail using midtime sampling methods, see Figs. \ref{fig_ssb_WASEP}(c)-(d) and Section \S\ref{s2c}. Although finite-size effects are strong as expected, a neat convergence towards the macroscopic MFT predictions is observed already for sizes $L=64$ when using a sufficiently large number of clones $N_c=2\times 10^4$. Similar finite-size effects appear in our measurements of the dynamical free energy $\mu(\lambda)$ and its derivatives, see Figs. \ref{fig_ssb_WASEP}(e)-(f), though again an excellent convergence to the MFT result is observed. Note in particular the non-quadratic shape of the dFE $\mu(\lambda)$ in the traveling-wave region, which corresponds to non-Gaussian current fluctuations in this regime \cite{perez-espigares13a}, as well as the second-order character of the transition, see $\mu''(\lambda)$ in Fig. \ref{fig_ssb_WASEP}(f).

\section{Dynamical criticality in high-dimensional driven systems}
\label{s4}

As we have already discussed in the introduction, the discovery of DPTs in the fluctuations of many-particle systems is an important finding for nonequilibrium physics. The reason is that the large deviation functions controlling the statistics of these fluctuations are the best candidates we have nowadays to generalize the concept of thermodynamic potentials to systems out of equilibrium, where no bottom-up approach exists yet connecting microscopic dynamics with macroscopic properties. In this way, understanding the singularities appearing in LDFs seems relevant to better understand nonequilibrium phenomena. In addition, the emergence of order associated with rare fluctuations implies that these extreme events are far more probable than previously anticipated \cite{lam09a,hurtado14a,zarfaty16a}. However, up to now most works on DPTs have focused on oversimplified $1d$ models as the ones discussed in previous section \cite{hurtado11a,perez-espigares13a,hurtado14a,vaikuntanathan14a,jack15a,shpielberg16a,zarfaty16a,baek17a,karevski17a,garrahan10a,ates12a,lesanovsky13a,carollo17a} or fluctuations of \emph{scalar} ($1d$) observables in $d>1$ \cite{garrahan07a,garrahan09a,hedges09a,chandler10a,pitard11a,speck12a,pinchaipat17a,abou18a,garrahan11a,genway12a,manzano14a}. The challenge thus remains to understand DPTs in more realistic settings relevant for actual experiments, as e.g DPTs in the fluctuations of fully vectorial observables in $d$ dimensions and how they are affected by the (possible) system anisotropy.

\begin{figure}
\vspace{-0.3cm}
\includegraphics[width=8.5cm]{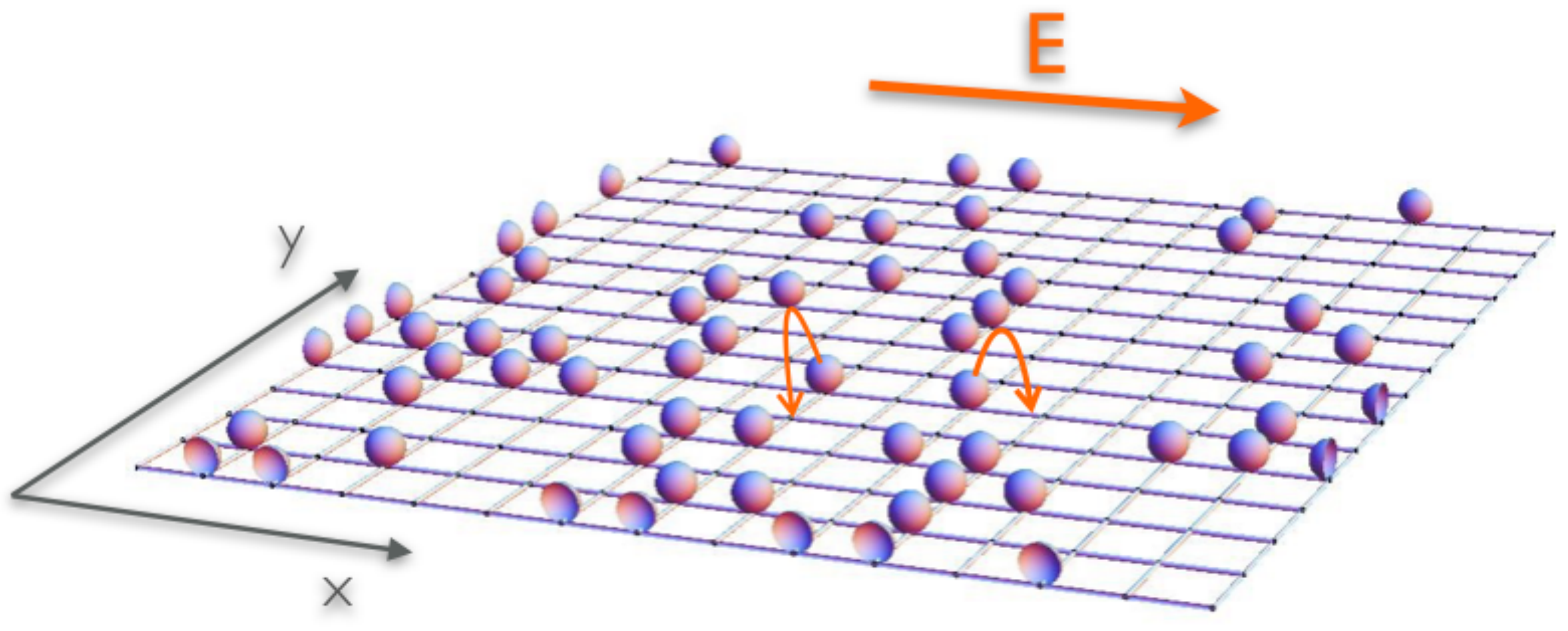}
\vspace{-0.2cm}
\caption{Sketch of the $2d$ WASEP. $M$ particles evolve in a $2d$ square lattice of size $N=L\times L$ with periodic boundary conditions.
Each site might be occupied by one particle at most, which jumps stochastically to neighboring empty sites at a rate $p^\alpha_{\pm}\equiv\text{exp}[\pm E_\alpha/L]/2$ for moves along the $\pm{\alpha}$-direction, $\alpha=x,y$, with $\vE=(E_x,E_y)$ the external vector field.
}
\label{fig4app}
\end{figure}

With this idea in mind, we employed the cloning Monte Carlo technique to study current fluctuations in a $2d$ version of the WASEP model presented in the previous section. In particular we define now the model on a $2d$ square lattice of size $N=L\times L$ with periodic boundaries (i.e. with the topology of a torus) where $M\leq N$ particles evolve, so the global density is $\rho_0 = M/N$. As before, each site may contain at most one particle. These particles can jump stochastically to neighboring empty sites, which now lie along the $\pm{\alpha}$-direction ($\alpha=x,y$), at a rate $p^\alpha_{\pm}\equiv\text{exp}[\pm E_\alpha/L]/2$, with $\vE=(E_x,E_y)$ being in this case a \emph{vectorial} external field. Fig. \ref{fig4app} shows an sketch of the $2d$ WASEP.

Interestingly, for large $\vE$ (needed to overcome a critical field value) and the moderate system sizes that we are capable to explore with the cloning algorithm, the field per unit length $\vE/L$ is strong enough to induce an \emph{effective anisotropy} in the medium. This effective anisotropy (a main feature of many real systems) enhances diffusivity and mobility along the strongest field direction, an effect that can be taken into account within macroscopic fluctuation theory via an effective anisotropy parameter $\epsilon$. 
Indeed, by expanding the microscopic transition rate $p_\pm ^\alpha$ to second order in the field per unit length, i.e. 
\be
p_\pm ^\alpha \approx \frac{1}{2}\left[1\pm \frac{E_\alpha}{L}+\frac{1}{2}\left(\frac{E_\alpha}{L}\right)^2\right] + {\cal O}\left[\left(\frac{E_\alpha}{L}\right)^3\right] \, , 
\ee
it is easy to show using a simple random walk argument that, while the first-order term gives rise to the standard field biasing dynamics along a given direction, the second-order perturbation results in effective differences of diffusivity and mobility along the different field directions.
In this way, the transport coefficients characterizing hydrodynamic behavior in the $2d$ WASEP are a diffusivity matrix $\hD(\rho)\equiv D(\rho)\mA$ and a mobility matrix $\hS(\rho)=\sigma(\rho)\mA$, with $D(\rho)=1/2$ and $\sigma(\rho)=\rho(1-\rho)$ as in the $1d$ WASEP of previous section, and  $\mA$ being a constant diagonal matrix 
\be
\mA = \left(\begin{array}{cc}
1+\epsilon & 0 \\
0 & 1-\epsilon
\end{array}
\right)
\label{Amatrix}
\ee
which modelizes 
the system underlying anisotropy in terms of an anisotropy parameter $\epsilon$.

\begin{figure}
\vspace{-0.3cm}
\includegraphics[width=8.5cm]{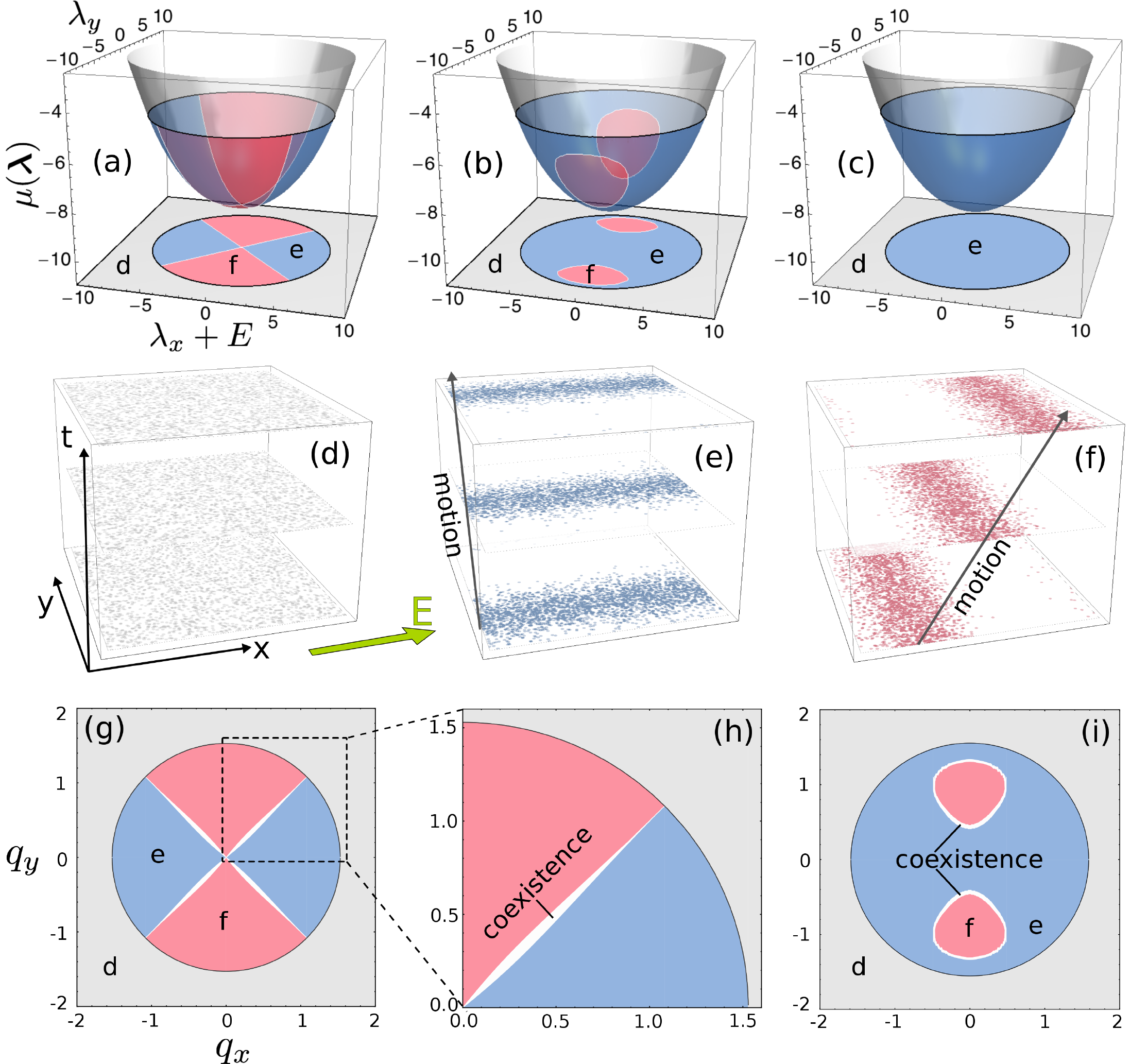}
\vspace{-0.2cm}
\caption{Top row: $\mu(\vlamb)$ for the $2d$-WASEP in an external field $\vE=(10,0)$, as derived from MFT, in the case of (a) no anisotropy, $\epsilon=0$, (b) mild anisotropy, $0<\epsilon<\epsilon_c$, and (c) strong anisotropy,  $\epsilon>\epsilon_c$. The projections show the phase diagram in $\vlamb$-space for each case, and letters indicate the typical spatiotemporal trajectories in each phase, displayed in the middle row (d)-(f). A DPT appears between a Gaussian phase (light gray) with homogeneous trajectories (d) and two different non-Gaussian symmetry-broken phases for low currents characterized by jammed density waves, (e) and (f). The first DPT is second-order, while the two symmetry-broken phases are separated by lines of first-order DPTs. Bottom row: phase diagram in current space for anisotropy $\epsilon=0$ (g,h), and $0<\epsilon<\epsilon_c$ (i). The coexistence pockets (white) are apparent. 
}
\label{fig1}
\end{figure}

The variational problem of MFT can be solved in this complex scenario, and predicts a rich dynamical phase diagram for the statistics of the space\&time-averaged current vector $\vq$ for the $2d$ WASEP, with some surprising new physics when compared to the $1d$ case. Fig. \ref{fig1} shows the predicted dynamical free energy $\mu(\vz)$, with $\vz\equiv \vlamb+\vE$, for different anisotropies $\epsilon$ as well as the dynamical phase diagram in each case and the typical trajectories in each dynamical phase. In particular, MFT predicts for large enough external fields $|\vE|>E_c$ a second-order DPT for currents $\vq\cdot\mA^{-1}\vq = \sigma_0^2\Xi_c$ (or equivalently $\vz\cdot\mA\vz= \Xi_c$), with $\sigma_0\equiv \sigma(\rho_0)$ and $\Xi_c$ a critical threshold. This critical DPT line separates a homogeneous fluctuation phase with structureless trajectories and Gaussian current statistics characterized by a quadratic dFE $\mu_{\text{G}}(\vz)\equiv (\vz\cdot \sigma_0\mA \vz - \vE\cdot\sigma_0\mA\vE)/2$, and a non-Gaussian dynamical phase for small currents, $\vq\cdot\mA^{-1}\vq \le \sigma_0^2\Xi_c$ or $\vz\cdot\mA\vz\le \Xi_c$. As in the $1d$ WASEP, coherent jammed states 
emerge in the non-Gaussian phase in the form of traveling-wave trajectories, thus breaking time-translational symmetry. Such jammed states hamper particle flow enhancing the probability of low-current fluctuations, but rather counter-intuitively these jammed states in $2d$ are surprisingly extended and noncompact. Interestingly, for mild or no anisotropy, $\epsilon<\epsilon_c$, different symmetry-broken $1d$ density waves dominate different current vector regimes, 
see Figs.~\ref{fig1}(a)-(b). Lines of first-order DPTs separate both density-wave phases, and dynamical coexistence emerges along these first-order lines between the two traveling-wave phases, see Figs. \ref{fig1}(g)-(i). However, the competition between the two different density-wave phases for low current fluctuations, modulated by the orientation of the current vector, disappears at a critical anisotropy $\epsilon_c\approx 0.035$, beyond which a single dynamical density-wave phase dominates the non-Gaussian regime, see Fig.~\ref{fig1}(c).

\begin{figure}
\vspace{-0.3cm}
\centerline{\includegraphics[width=9cm]{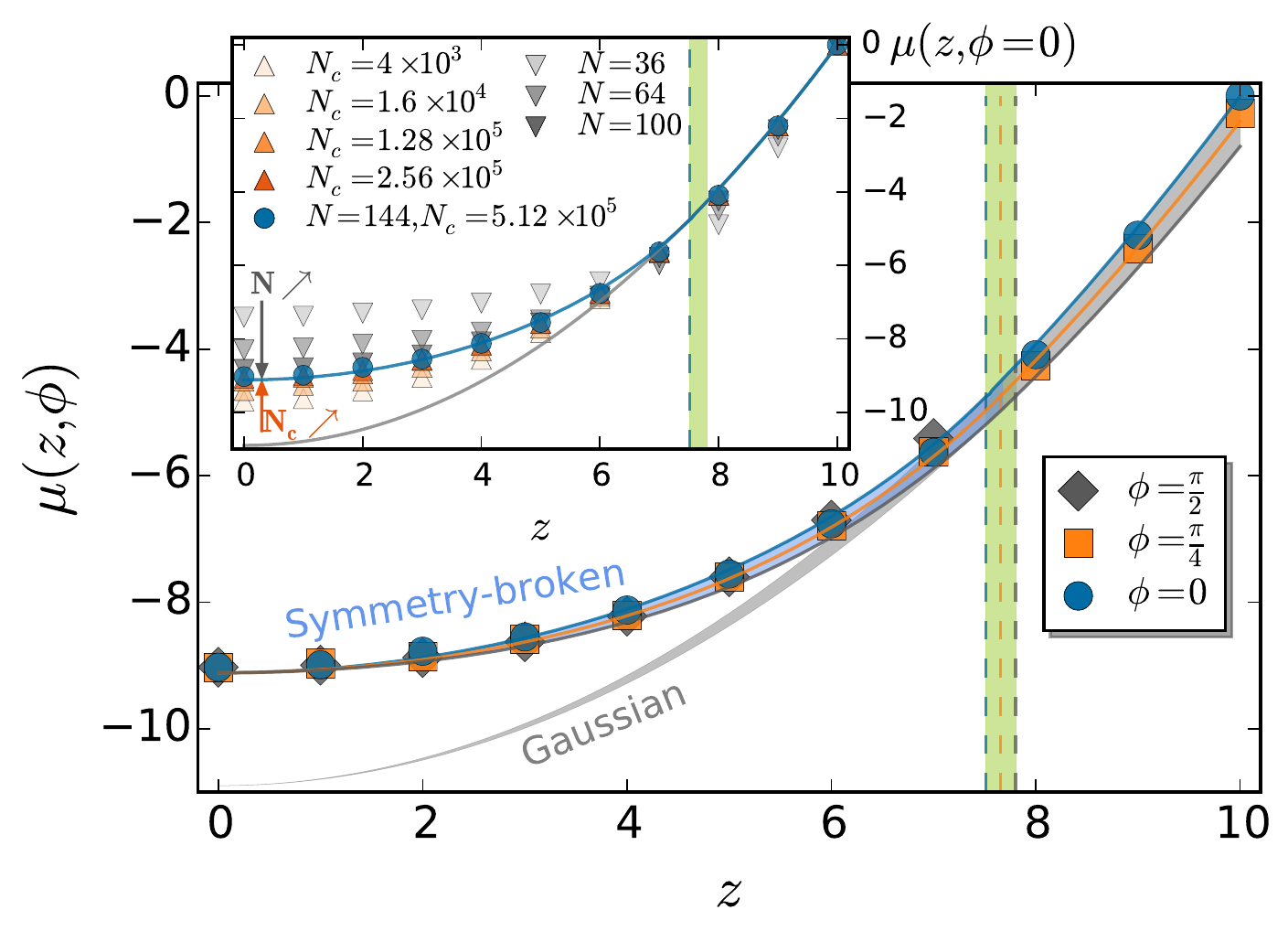}}
\vspace{-0.4cm}
\caption{Main: $\mu(\vlamb)$ vs $z=|\vlamb+\vE|$ as obtained in simulations for $N=144$, $N_c=5.12\times10^5$ and different $\phi=\tan^{-1}(z_y/z_x)$, together with MFT predictions for anisotropy $\epsilon=0.038$. A DPT from a Gaussian regime (light-gray ribbon) to a symmetry-broken, non-Gaussian phase (blue ribbon) is apparent upon crossing $z_c(\phi)$, with $\vz_c\cdot\mA\vz_c=\Xi_c$ (green vertical stripe). Different $\phi$ correspond to different MFT lines within the shaded ribbons. Inset: Convergence to the $\phi=0$ MFT prediction (blue line) for $N=144$ as $N_c$ increases ($\bigtriangleup$) and for optimal $N_c$ as $N$ increases  ($\bigtriangledown$). 
}
\label{fig2}
\end{figure}

Our aim here is to search for compelling evidences of this complex DPT in numerical simulations of the $2d$ WASEP, and to characterize in detail the emergent order predicted for low currents. To do so, we explored the statistics of the particle vector current in this model using massive cloning Monte Carlo simulations \cite{giardina06a,lecomte07a,giardina11a,tizon-escamilla17b}, see Section \S\ref{s2}. In particular, we focused on density $\rho_0=0.3$ under a strong external field $\vE=(10,0)$ along the $\hat{x}$-direction, such that $|\vE|>E_c$, investigating in detail finite-size effects up to sizes $N=144$. Moreover, as the number of clones needed to observe a given rare event typically grows exponentially with the system size, all the more the rarer the event is \cite{hurtado09a,nemoto16a}, we needed to reach the extraordinary number of $N_c=5.12\times10^5$ clones evolving in parallel for a long time to correctly sample the tails of the current distribution. 

The main panel in Fig.~\ref{fig2} shows the measured $\mu(\vz)$ for $N=144$ and $N_c=5.12\times10^5$ as a function of $z=|\vz|$ for different current orientations $\phi=\tan^{-1}(z_y/z_x)$, while the inset analyzes both finite-$N$ and finite-$N_c$ corrections in our measurements as a function of $z$ for $\phi=0$. In particular, the inset shows the effect of a varying number of clones on the measured $\mu(\vz)$ for the largest $N=144$: while finite-$N_c$ corrections are weak for $z\approx E$ (i.e. $\vlamb \approx 0$ or $\vq\approx \vq_{st}= \sigma_0\mA \vE$, small fluctuation regime),
as otherwise expected, these corrections mount up as $z\to 0$ (equivalently $\vlamb \to -\vE$ or $\vq\to 0$)
tending to overestimate the value of $|\mu(\vz)|$.  In any case, these corrections scale as $1/N_c$, in agreement with \cite{guevara17a,nemoto17a}, so as to observe an excellent convergence toward the macroscopic fluctuation theory prediction (solid line) for large enough $N_c$ and the largest $N$ explored, as displayed in the inset to Fig.~\ref{fig2}. On the other hand, finite-$N$ corrections measured for the largest $N_c=5.12\times10^5$ affect equally our measurements of the dFE irrespective of the value of $z$, reflecting the \emph{distance} of the microscopic simulation to the predicted behavior based on a macroscopic approach. These finite-$N$ corrections, which underestimate $|\mu(\vz)|$ for large fluctuations ($z\to 0$), decay at good pace as $N$ increases, showing an excellent convergence to the MFT prediction already for $N=144$ and $N_c=5.12\times10^5$.

\begin{figure}
\vspace{-0.3cm}
\centerline{\includegraphics[width=8.5cm]{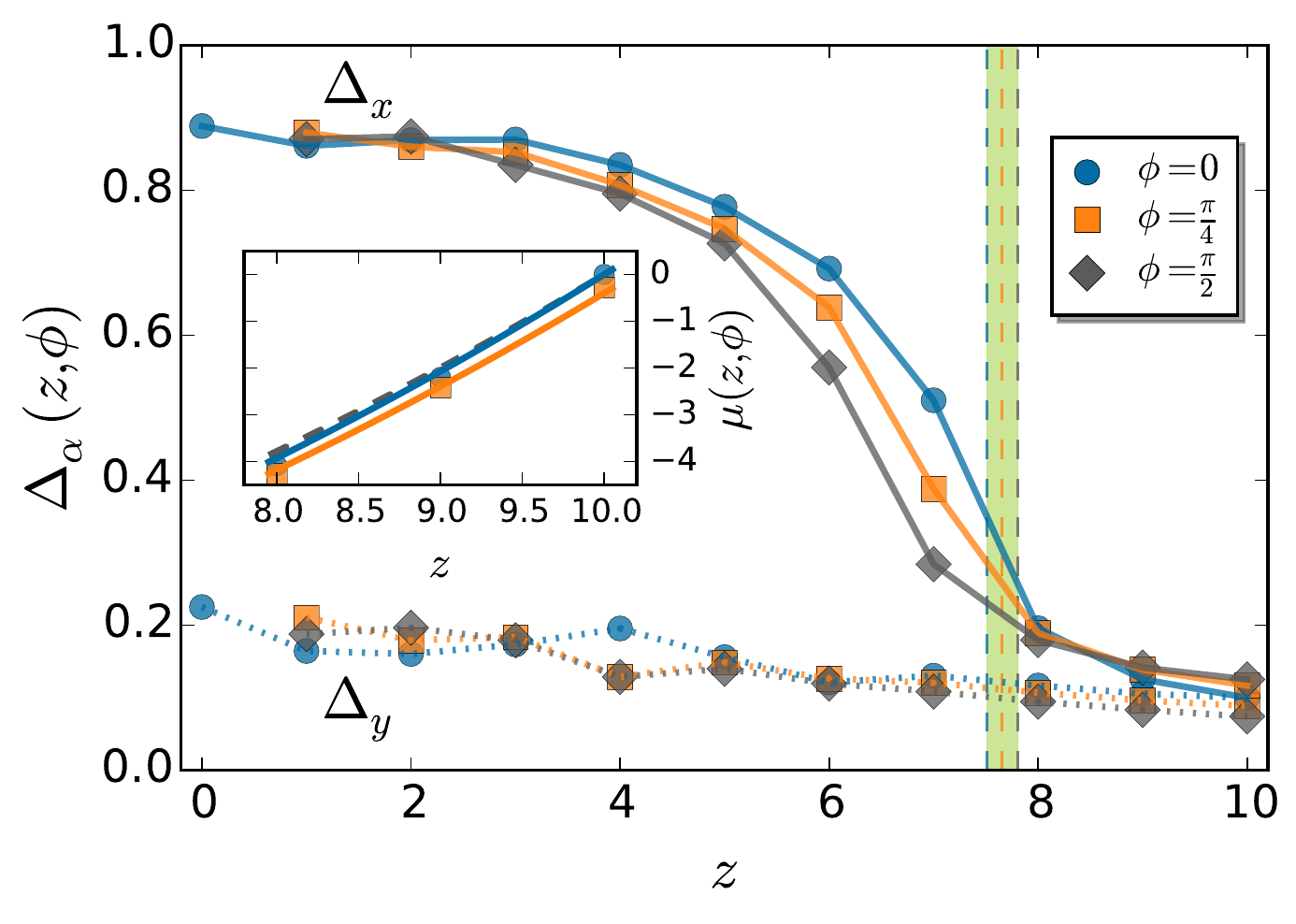}}
\vspace{-0.5cm}
\caption{Tomographic $\alpha$-coherences, with $\alpha=x,y$, as a function of $z$ for different current angles $\phi$ measured for $N=100$ and $\vE=(10,0)$. Inset: dFE $\mu(\vz)$ vs $z$ in the Gaussian regime for $\phi=0,\pi/4$, see Fig.~\ref{fig2}. Full (dashed) lines are MFT predictions with anisotropy $\epsilon=0.038$ ($\epsilon=0$).}
\label{fig3}
\end{figure}

In this way, and in agreement with MFT, the measured dFE is fully compatible with the Gaussian prediction $\mu_{\text{G}}(\vz)$ for $\vz\cdot\mA\vz\ge \Xi_c$. This confirms the idea that small current fluctuations come out from the random superposition of typically uncorrelated, localized jump events which sum up incoherently to yield Gaussian statistics and structureless typical trajectories (see below). 
Interestingly, our data exhibit a weak dependence of the dFE $\mu(\vz)$ on the angle $\phi$ in this Gaussian regime, a clear hallmark of the effective anisotropy $\epsilon$ mentioned above. Indeed, by fitting the observed data in the region $\vz\cdot\mA\vz\ge \Xi_c$ to the $\epsilon$-dependent quadratic form $\mu_{\text{G}}(\vz)\equiv (\vz\cdot \sigma_0\mA \vz - \vE\cdot\sigma_0\mA\vE)/2$ for each angle $\phi$, see inset in Fig.~\ref{fig3}, we can estimate the effective value of the anisotropy parameter for $N=144$, obtaining that $\epsilon\approx 0.038$ properly describes the observed weak anisotropy.
This effective anisotropy is slightly larger than the critical anisotropy $\epsilon_c\approx 0.035$ beyond which a single symmetry-broken phase dominates the non-Gaussian regime, see Fig.~\ref{fig1}(c), an observation consistent with additional results at the trajectory level (see below). Furthermore, Fig.~\ref{fig2} clearly shows that the Gaussian fluctuation regime ends up for $\vz\cdot\mA\vz<\Xi_c$, where systematic deviations from the quadratic form $\mu_{\text{G}}(\vz)$ become apparent. This change of behavior, in excellent agreement with MFT predictions, signals the onset of the DPT into a non-Gaussian current fluctuation regime characterized by traveling density-wave trajectories.

\begin{figure*}
\vspace{-0.3cm}
\includegraphics[width=14.5cm]{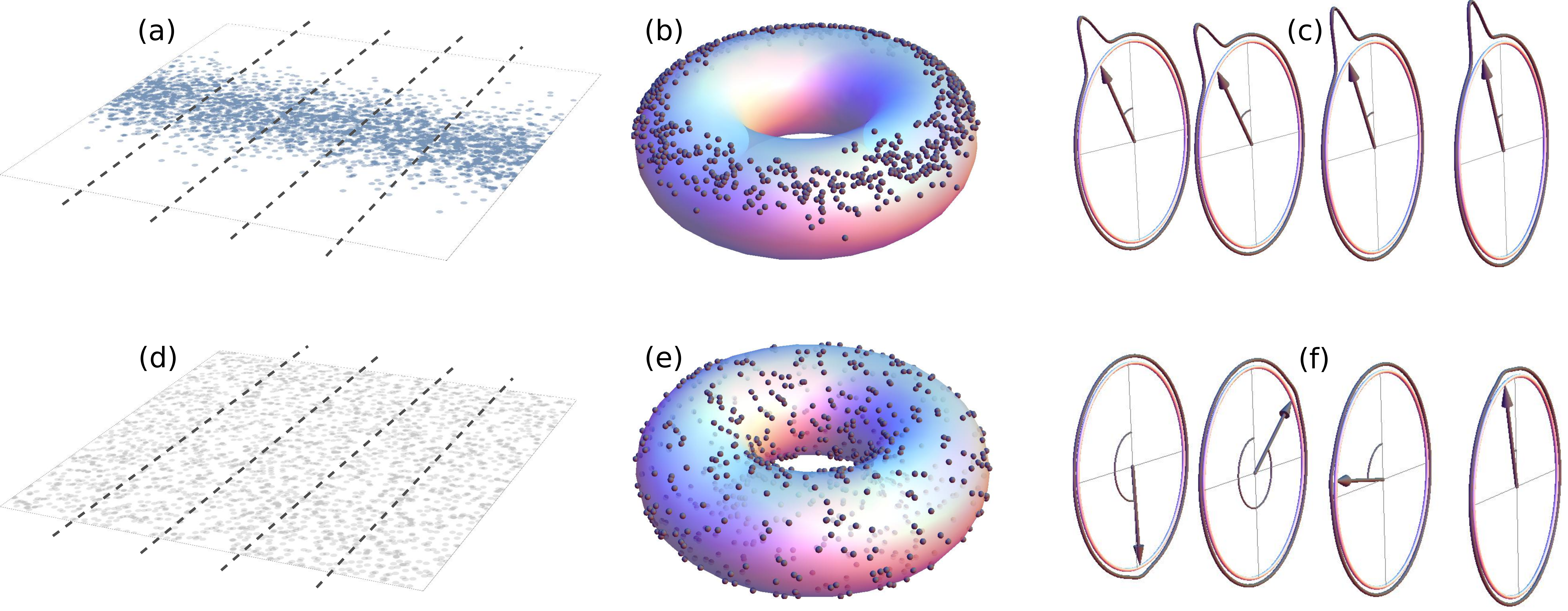}
\vspace{-0.2cm}
\caption{Tomographic analysis to define an order parameter for the DPT in the $2d$ WASEP. Order is expected to emerge across the DPT in the form of $1d$ coherent traveling waves (a) which jam particle flow along one direction. To detect these jams, we slice microscopic configurations along principal axes (see dashed lines in (a)). Due to the periodic boundaries, the system topology is in fact that of a torus, as in (b), so each slice can be considered as a $1d$ ring of fixed radius embedded in a two-dimensional space, with a given angular mass distribution (c) depending on the positions of the particles in the slice. A small dispersion $\Sigma_x^2$ of the angular centers of mass across the different slices, (c), will signal the formation of a coherent jam along the $x$-direction and the associated density wave in the orthogonal direction, see (a). A similar analysis in the homogeneous, Gaussian phase leads to a typically large dispersion $\Sigma_x^2$, see (d)-(f).
}
\label{fig5app}
\end{figure*}

The observed crossover in the dFE at $\vz\cdot\mA\vz=\Xi_c$, see Fig.~\ref{fig2}, is a strong evidence of the existence of the predicted DPT.
In order to confirm this evidence, and to fully characterize the DPT, we now search for an order parameter exhibiting a smooth but apparent change as we cross the critical line. This is the smoking gun of any continuous phase transition such as the DPT here studied \cite{binney92a}. Moreover, the order parameter introduced must distinguish between the competing symmetry-broken phases  (i.e. the different jammed density-wave trajectories) which are expected to appear for low current fluctuations. Such order parameter can be defined by performing a \emph{tomographic analysis} of configurations, i.e. by taking $1d$ slices of our $2d$ system and analyzing density profiles in each slice. In particular we consider a microscopic particle configuration ${\bf n}=\{n_{ij};i,j\in [1,L] \}$ and slice it along one of the principal axes, say $\hat{x}$, defining the configuration of the $j$-slice as 
\be
{\bf n}_j\equiv\{n_{ij};i\in [1,L] \} \, , \nonumber
\ee
see e.g Figs. \ref{fig5app}(a) and \ref{fig5app}(d). To properly take into account the periodic boundaries [i.e. the system torus topology, see Fig.~\ref{fig5app}(b) and Fig.\ref{fig5app}(e)], and in analogy with the angular analysis introduced in Section \S\ref{s3a} for the $1d$ case, we consider each $j$-slice as a ring of fixed radius embedded in $2d$, so we can assign to each site $i\in[1,L]$ an angle $\theta_i=2\pi i/L$, and compute the angular position of the center of mass for the $j$-slice, $\theta_{\text{cm}}^{(j)}$. This is defined as
\be
\theta_{\text{cm}}^{(j)}\equiv \tan^{-1}\left(\frac{S_j}{C_j}\right)
\ee
where we have introduced the definitions
\ben
S_j &\equiv& \frac{1}{M_j}\sum_{i=1}^Ln_{ij}\sin\theta_i \, , \\
C_j &\equiv& \frac{1}{M_j}\sum_{i=1}^Ln_{ij}\cos\theta_i \, ,
\een
with $M_j=\sum_{i=1}^{L}n_{ij}$ the total number of particles in this slice. In this way, a small dispersion of the angular centers of mass across the different slices clearly signals the formation of a coherent jam along the $x$-direction and the associated density wave in the orthogonal direction, see Fig. \ref{fig5app}(c). On the other hand, a large dispersion of $\theta_{\text{cm}}^{(j)}$ across the different $j\in[1,L]$ is the typical signature of a structureless, homogeneous random configuration, see Fig.~\ref{fig5app}(d) and Fig.\ref{fig5app}(f). Therefore we introduce a measure of such dispersion as the variance of the angular centers of mass across the different slices,
\be
\Sigma_x^2\equiv \la(\theta_{\text{cm}}^{(j)})^2\ra_x - \la{\theta_{\text{cm}}^{(j)}}\ra_x^2 \, ,
\ee
where we have defined
\be
\la f_j\ra_x \equiv \frac{1}{L}\sum_{j=1}^L f_j \, ,
\ee
for any arbitrary local observable $f_j$. We finally define the \emph{tomographic $x$-coherence} as 
\be
\Delta_x(\vlamb)\equiv 1-\la\Sigma_x^2\ra_\vlamb \, ,
\ee
where the average $\la\cdot\ra_\vlamb$ is taken over the biased $\vlamb$-ensemble, i.e. over all trajectories statistically relevant \emph{during} a rare event of fixed current-conjugated parameter $\vlamb$. We can define in an equivalent way the tomographic $y$-coherence $\Delta_y(\vlamb)$ to detect particle jams along the $y$-direction, and Fig.~\ref{fig3} shows these two order parameters measured across the DPT as a function of $z= |\vlamb+\vE|$. 

The results obtained with these two complementary order parameters fully confirm the presence of the DPT, shedding additional light on the symmetry-breaking process. In particular, $\Delta_x(z)$ increases steeply for $\vz\cdot\mA\vz\le \Xi_c$ and \emph{all angles} $\phi$ of the current vector, while $\Delta_y(z)$ remains small and does not change appreciably across the DPT. This phenomenology clearly indicates that a coherent particle jam emerges along the $x$-direction for $\vz\cdot\mA\vz\le \Xi_c$ and all angles, as in the sketch of Fig. \ref{fig5app}(a) above. This means that only one of the two possible symmetry-broken density-wave phases appear in our simulations (regardless of the current vector orientation). Such result agrees with the MFT prediction in the supercritical anisotropy regime $\epsilon>\epsilon_c$, see Fig.~\ref{fig1}(c), and is consistent with the measured effective anisotropy $\epsilon\approx0.038>\epsilon_c$, see inset in Fig.~\ref{fig3}. In addition, the behavior of both tomographic coherences $\Delta_\alpha$ ($\alpha=x,y$) across the DPT is consistent with the emergence of a traveling density-wave with structure in $1d$ and not in $2d$, as in the latter case both $\Delta_\alpha$ should increase upon crossing the critical line. Moreover, the acute but continuous change of $\Delta_x(\vz)$ across the DPT is consistent with a second-order transition, in agreement with the MFT prediction.

\section{Particle-hole symmetry breaking at the fluctuation level}
\label{s5}

The dynamical phase transitions in current statistics reported so far take place in periodic settings. In such systems, the additivity principle (which conjectures the time-independence of the dominant trajectories responsible of a fluctuation, see Section \S\ref{s2}) is violated via the emergence of traveling-wave density profiles that break the time-translational invariance. For several years it was not clear whether other types of DPTs could appear for \emph{open} diffusive systems, where energy or particles are injected through reservoirs attached to the boundaries. However, it was recently shown that first- and second-order DPTs might also occur for open systems with an external field through a novel mechanism: particle-hole (PH) symmetry breaking \cite{baek17a,baek18a}. This important prediction, based on a perturbative Landau theory derived within MFT, is however restricted to (a) fluctuations around the critical current and (b) equal boundary densities/temperatures (or at most infinitesimally small gradients). Additionally, its microscopic origin is not understood yet and, most importantly, such a DPT has never been observed in numerical experiments, which might offer clues on novel phenomenology far from the critical point not yet explored. To shed light on all these issues, we have thoroughly studied the open $1d$ WASEP using both MFT and cloning Monte Carlo simulations in search of this elusive DPT \cite{perez-espigares18c}. 

\begin{figure}
\vspace{-0.2cm}\includegraphics[scale=0.28]{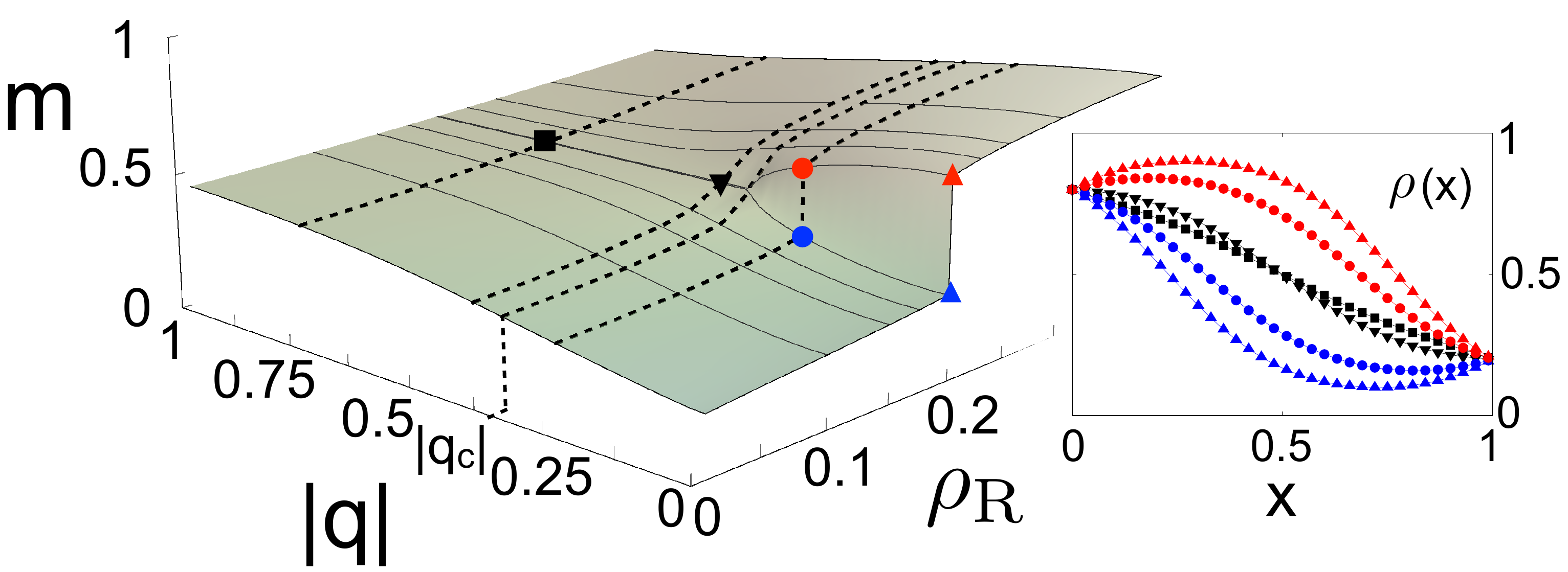}
\vspace{-0.4cm}
\caption{
Mass $m_q$ of the optimal trajectory responsible for a current fluctuation $q$ for different boundary drivings, 
with $\rho_{\text{L}}=0.8$, $\rho_{\text{R}}\in[0,0.4]$ and external field $E=4$. Inset: Optimal profiles derived from the MFT for $\rho_{\text{R}}=0.2$ and $q$'s signaled in the main plot. 
}\label{fig1oWASEP}
\end{figure}

The open $1d$ WASEP, already described in Section \S\ref{s3b}, is characterized at the microscopic level by biased bulk jump rates $p^{\pm}=pe^{\pm E/L}$ to the right ($+$) and to the left ($-$), see Fig.~\ref{figmodelSSEP}(a).
Taking $p=1/2$ for simplicity, this model is characterized at the macroscopic level by a diffusivity $D(\rho)=1/2$ and a mobility $\sigma(\rho)=\rho(1-\rho)$. For these transport coefficients, the current large deviation function $G(q)$ --given by Eq. \eqref{LDFap}-- remains invariant under a PH transformation, namely $\rho \to 1-\rho$ and $x \to 1-x$, whenever the densities at the boundaries satisfy the condition $\rho_\text{L}=1-\rho_\text{R}$. Interestingly, the work in \cite{baek17a,baek18a} predicts that this PH symmetry can be eventually broken when the system is conditioned to maintain atypically low currents in the presence of a large enough field ($|E|>E_c=\pi$). 
More in detail, in order to sustain a current $q$ above a critical threshold $(|q|>|q_c|)$, the system adopts an optimal profile $\rho_q(x)$ that is PH-symmetric, i.e. $\rho_{q}(x)=1-\rho_{q}(1-x)$, so the optimal profile inherits the symmetry of the \emph{governing action}, the current LDF \eqref{LDFap}. This is illustrated by the black density profiles of the inset to Fig.~\ref{fig1oWASEP}, corresponding to the currents given by the black symbols in the main panel. Nevertheless, MFT predicts that for currents below the critical point $(|q|<|q_c|)$ the PH symmetry of the current LDF is broken: two coexisting symmetry-broken optimal profiles $\rho_{q}^{\pm}(x)$ appear, such that $\rho_{q}^{\pm}(x)\neq 1-\rho_{q}^{\pm}(1-x)$. 
These two different profiles are however linked by the PH transformation, i.e. $\rho_{q}^{\pm}(x)=1-\rho_{q}^{\mp}(1-x)$, thus restoring the broken symmetry.  This is displayed in the inset to Fig.~\ref{fig1oWASEP} by the red, $\rho_{q}^+(x)$, and blue profiles, $\rho_{q}^-(x)$, which are respectively associated with the currents given by the red and blue symbols of the main panel. As argued in \cite{perez-espigares18c}, this symmetry-breaking phenomenon is captured by a global order parameter such as the total mass of the system $m=\int_0^1 \rho(x) dx$. In particular, by studying the joint fluctuations of the current $q$ and this collective order parameter $m$, the dynamical phase diagram for arbitrary boundary gradients (both symmetric and asymmetric) can be unveiled. This is shown in the main panel of Fig.~\ref{fig1oWASEP}, which displays the mass $m_q$ of the optimal trajectory responsible for a current fluctuation $q$ for different boundary drivings. Note the $\mathbb{Z}_2$-type symmetry-breaking transition appearing at a critical current $|q_c|$ for $\rho_\text{R}=1-\rho_\text{L}$.

\begin{figure*}
\includegraphics[width=18cm]{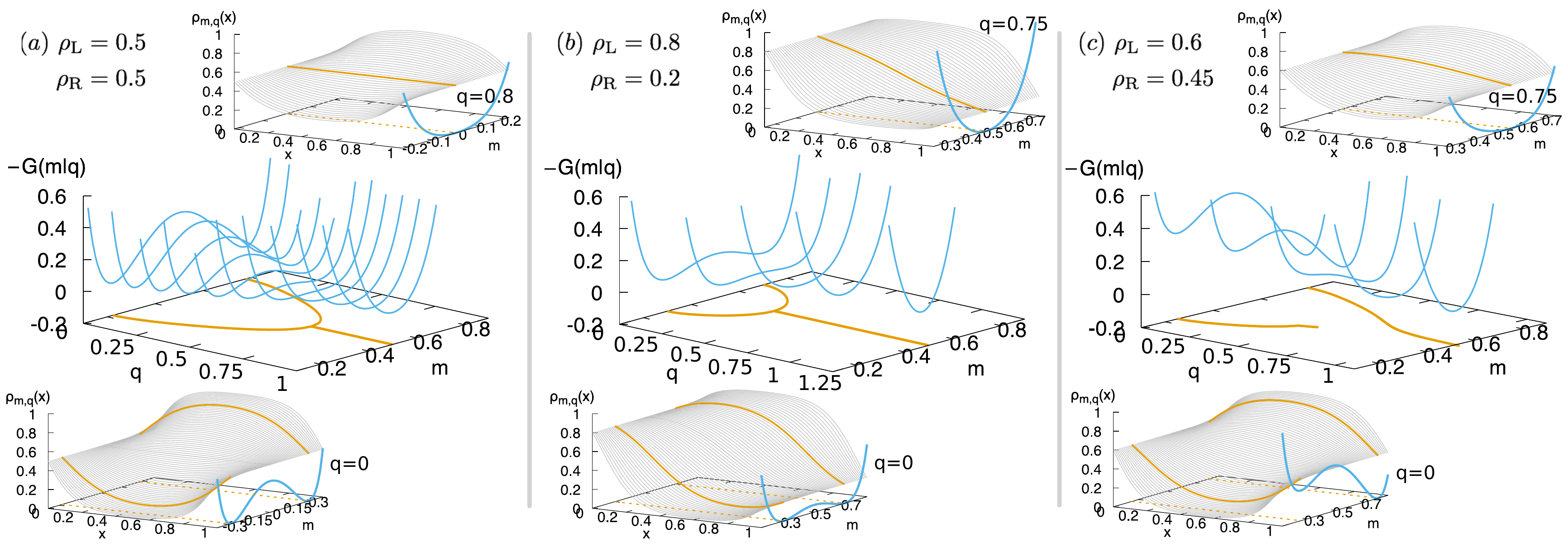}
\caption{Middle row: (Negative) Conditional LDF $-G(m|q)=G(q)-G(m,q)$ as a function of the mass $m$ for different currents $q$ for $E=4$ and three different boundary drivings, namely (a) $\rho_{\text{L}}=0.5, \rho_{\text{R}}=0.5$ (symmetric driving), (b) $\rho_{\text{L}}=0.8, \rho_{\text{R}}=0.2$ (symmetric driving), and (c) $\rho_{\text{L}}=0.6, \rho_{\text{R}}=0.45$ (asymmetric driving). The lines projected in the $m-q$ plane correspond to the local minima of the negative LDF, $-G(m|q)$, which define the mass $m_q$ associated with a current fluctuation $q$. In the symmetry-broken regime this defines the low- and high-mass branches  $m_q^\pm$. Bottom row: optimal density profiles $\rho_{m,q}(x)$ obtained from the MFT for $q=0$ and the three different boundary drivings. The thick lines are the optimal profiles associated with the local minima $m_q^\pm$ of $-G(m|q)$. For completeness the corresponding $-G(m|q)$ is also shown. Top row: optimal  MFT density profiles in each case, for a current in the PH-symmetric region, $|q|>q_c$.}
\label{fig1SM}
\end{figure*}

The DPT here reported for the open WASEP is, as its periodic counterpart (see Section \ref{s3b}), of second-order type and takes place whenever the densities at the boundaries are PH symmetric, $\rho_\text{L}=1-\rho_\text{R}$. As in standard critical phenomena, much information of the transition can be obtained by studying fluctuations of the order parameter (i.e. the total mass) across the critical point. In this case this means investigating the
the joint mass-current LDF $G(m,q)$, or rather the conditional LDF $G(m|q)=G(m,q)-G(q)$. This can be computed from \eqref{LDFap} just by minimizing over density profiles with the additional constraint $m=\int_0^1 \rho(x) dx$. Fig.~\ref{fig1SM} illustrates the MFT prediction for $-G(m|q)$ and different boundary drivings. Note in particular that the optimal mass $m_q$ for a current fluctuation $q$, shown in the main panel of Fig.~\ref{fig1oWASEP}, is nothing but the mass
minimizing $-G(m|q)$ for each $q$. 
Fig.~\ref{fig1SM} also shows the optimal profiles $\rho_{m,q}(x)$, derived from the MFT, associated with each mass $m$ for different currents $q$ (top and bottom rows) and different boundary drivings with $E=4$. For $\rho_\text{L}=1-\rho_\text{R}$, the joint LDF $-G(m|q)$ moves from a region with one global minimum to a symmetry-broken phase with two global minima as $|q|$ crosses a critical threshold $|q_c|$, a behavior typical of a $\mathbb{Z}_2$-type transition, and this is accompanied by an emergent degeneration in the optimal profiles for $|q|<|q_c|$, $\rho_{q}^{\pm}(x)=1-\rho_{q}^{\mp}(1-x)$, see the top and bottom rows of Figs.~\ref{fig1SM}(a)-(b).
On the other hand, for asymmetric boundary drivings, $\rho_\text{L}\neq 1-\rho_\text{R}$, there is always one single global minimum in $-G(m|q)$ $\forall q$, although a local minimum might appear for low enough currents associated with dynamical metastability, see Fig.~\ref{fig1SM}(c).

A natural question that arises is whether a time-dependent density profile might become a better minimizer of the LDF $G(m|q)$ for some values of $m$. Interestingly, the answer is affirmative in the symmetry-broken phase, as a result of the non-convexity of $G(m|q)$ \cite{perez-espigares18c}. In this case, a \emph{dynamical time-coexistence} between the upper- and lower-branch optimal profiles $\rho_q^\pm(x)$ gives rise to the convex envelope of $G(m|q)$ for $m\in (m_q^-,m_q^+)$, with $m_q^{\pm}=\int_0^1 \rho_{q}^{\pm}(x) dx$. This means that in order to produce a mass $m\in (m_q^-,m_q^+)$ sustaining a current $q$ during a long time interval $\tau$, the system adopts optimal profile is $\rho_{q}^{+}(x)$ for a fraction of time $\nu \tau$ with $\nu \in [0,1]$ while it maintains $\rho_{q}^{-}(x)$ for the rest of the time $(1-\nu)\tau$. This coexistence in time results in a convex envelope --as a Maxwell-like construction-- which reads $G(m|q)=\nu G(m_q^{-}|q)+(1-\nu)G(m_q^+|q)$, thus maximizing the probability $P(m|q)$, as in a standard first-order phase transition \cite{perez-espigares18c}.

\begin{figure}
\vspace{0cm}\includegraphics[scale=0.245]{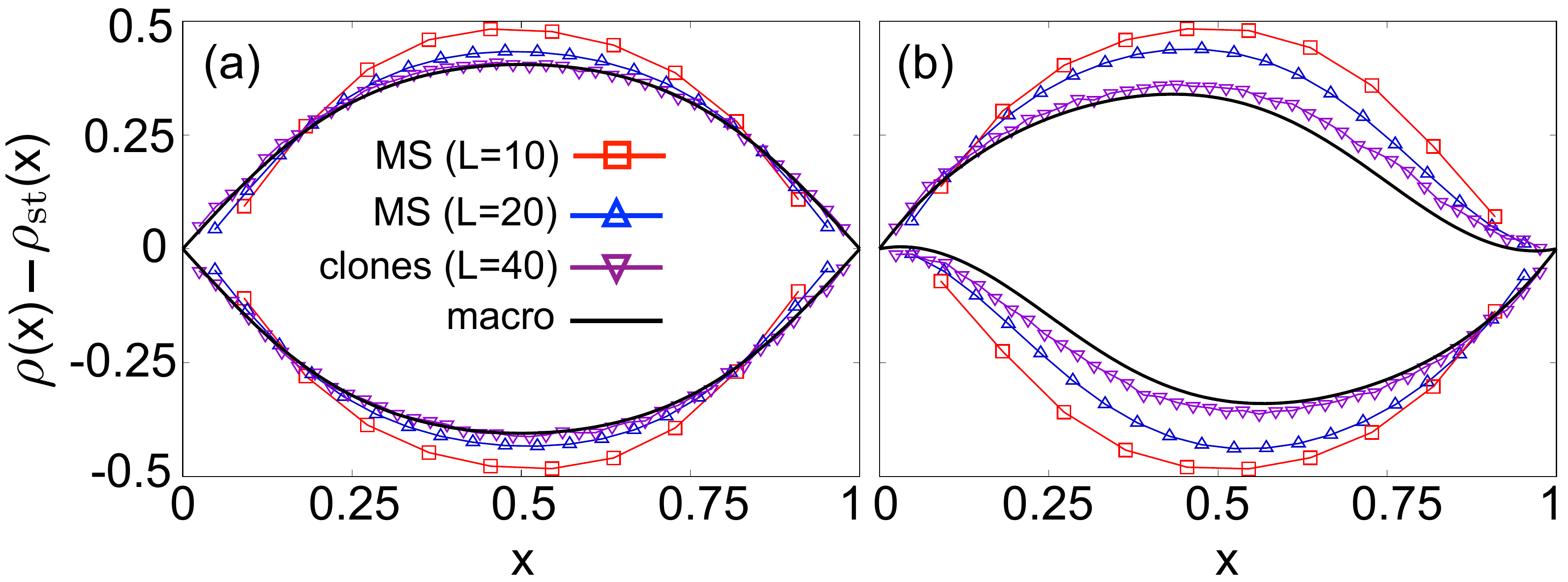}
\vspace{-0.4cm}
\caption{
(a) Optimal density profiles for the open $1d$ WASEP with $\rho_{\text{L}}=\rho_{\text{R}}=0.5$ and $E=4$ associated with $q=0$. Macroscopic predictions (black solid lines) and simulation results using the cloning algorithm for $L=40$ (purple down triangles). Also shown are density profiles associated with the extremal metastable states for $L=10$ (red squares) and $L=20$ (blue up triangles). (b) Same results for  $\rho_{\text{L}}=0.8$ and $\rho_{\text{R}}=0.2$. 
}\label{fig4}
\end{figure}

So far we have analyzed the DPT at a macroscopic scale, however, in order to have a better understanding of the phenomenon we set out to study its origin from the microscopic dynamics. This approach may be carried out via exact diagonalization of the generator of the dynamics and by means of cloning numerical simulations. The former analysis shows that the DPT here reported corresponds to a degeneracy of the ground state (the one associated with the leading eigenvalue) of the dynamical generator. It turns out that its spectral gap closes as $L$ increases, i.e. the sub-leading eigenvalue coalesces with the leading one. For large but finite $L$, this gap is small but non-zero, thus giving rise to two long-lived metastable states (MS) which converge to the macroscopic symmetry-broken profiles for increasing sizes [see density profiles with square and triangle symbols in Figs.~\ref{fig4}(a)-(b)]. Details on the derivation of these MS optimal profiles, obtained using the so-called Doob's h-transform, can be found in Ref.\cite{perez-espigares18c}. 

Despite the physical understanding resulting from the spectral analysis, a direct observation of this phenomenon --occurring in the far tails of the current distribution--, remains to be done. For that reason we have carried out numerical simulations using the cloning Monte Carlo algorithm described in previous sections. This method allows us to reach larger system sizes than with the diagonalization method and to confirm the theoretical predictions for different boundary drivings.
Results are shown in Figs.~\ref{fig4}(a)-(b), where the optimal profiles in the symmetry broken phase (associated with $q=0$, thus corresponding to $\lambda=-E$ with $E=4$) have been displayed with purple down triangles both for (a) $\rho_\text{L}=\rho_\text{R}=0.5$ and (b) $\rho_\text{L}=0.8$ and $\rho_\text{R}=0.2$, using a population of $N_c=10^4$ clones. It is worth noting that, in order to correctly average the optimal profiles in the cloning simulations, we have to distinguish between those realizations sustaining a mass \emph{above} $1/2$ and those sustaining a mass \emph{below} $1/2$. Thus, averaging them separately, in order not to blur away the correct spatial structure of the density profiles, we obtain the results depicted in Fig.~\ref{fig4}. Taking together the density profiles obtained from the spectral analysis of the microscopic generator and those obtained from the cloning Monte Carlo algorithm, we observe a clear convergence toward the theoretical MFT (symmetry-broken) prediction, though finite-size effects are still present. This suggest to pursue further cloning simulations introducing the recent advances of Section \S\ref{s2e} to minimize sampling issues so as to reach larger system sizes with a reasonable number of clones. This will be the focus of future work.

\section{Conclusions} 
\label{s6}

In this paper we have reviewed a powerful Monte Carlo technique to sample rare events in many-particle systems. 
This computational method consists in modifying the system dynamics so that the rare events of interest become no longer rare, and involve the simultaneous evolution of multiple copies or \emph{clones} of the system, which replicate or die in time according to their statistical weight. In particular, we have described in detail two different versions of this general method, namely as applied to discrete- and continuous-time stochastic many-particle systems. Special emphasis has been put on understanding the role of finite-size effects (both on the number of clones and the system size) on the sampling of the rare events of interest, as well as on the measurement and characterization of average trajectory observables characterizing a given rare event (midtime statistics). 

The application of this new computational tool to simple models, particularly stochastic lattice gases, provides intriguing evidences of the existence of rich and fundamental structures in the fluctuating behavior of nonequilibrium systems, which typically emerge via dynamical phase transitions (DPTs).  These DPTs, which are the focus of the second part of this work, appear when a system with many degrees of freedom sustains atypical fluctuations of dynamical observables such as the current or the activity. This leads in some cases to symmetry-broken space-time trajectories which enhance the probability of such events due to the emergence of ordered structures. 

Despite their importance, and due to their low probability of occurrence, these DPTs are extremely difficult to characterize empirically. This makes the cloning Monte Carlo method an ideal tool to observe and characterize for the first time complex DPTs in many-particle systems. In this way, we have described some tricks of the trade related to the sampling of rare events across dynamical phase transition using the cloning method, with an emphasis on the definition and measurement of order parameters capturing the physics of the different DPTs, and the characterization of the optimal paths responsible for a fluctuation. In particular, we have described the application of the cloning method to uncover different DPTs in the current statistics of two paradigmatic models of transport, namely the Kipnis-Marchioro-Presutti (KMP) model of heat transport and the weakly asymmetric simple exclusion process (WASEP). These models exhibit spontaneous breaking of time-translation symmetry at the trajectory level under periodic boundary conditions, via the appearance of a time-dependent traveling wave. Despite their mathematical similarities, these DPTs have a radically different interpretation: while the DPT in the KMP model appears for large enough currents, where ballistic energy packets emerge to facilitate these fluctuations, in the WASEP the transition kicks in for low currents, where jammed density-wave trajectories dominate. 

These DPTs are fully confirmed in numerical cloning experiments, which allow a detailed characterization of the emergence of order across the transition. Moreover, the versatility of the cloning Monte Carlo method has allowed us to explore the role of dimensionality in DPTs. In particular, we have investigated the \emph{vector} current statistics in the $2d$ WASEP, where the complex interplay among an external field, the possible system anisotropy, and the vectorial character of currents leads to a rich phase diagram at the fluctuating level, with different symmetry-broken fluctuation phases separated by lines of first- and second-order DPTs. These predictions, based on macroscopic fluctuation theory, are fully confirmed in numerical cloning experiments by introducing novel order parameters to characterize the competing ordered phases which emerge for low currents. Finally, we have explored both theoretically and numerically a different type of symmetry-breaking phenomenon at the trajectory level which appears in \emph{open} systems, i.e. coupled to boundary reservoirs which may impose an external gradient (of e.g. density or temperature). In this case a novel DPT appears which breaks the particle-hole symmetry --a $\mathbb{Z}_2$ discrete symmetry--, a transition that persists in the presence of arbitrarily strong (but symmetric) boundary gradients. The cloning method allows to observe and characterize for the first time this transition.

While the DPTs studied here appear in transport models of increasing complexity, we are still far from a complete understanding of the possible dynamical phases which may emerge in the fluctuations of more realistic systems, as e.g. hydrodynamic-type media characterized by several coupled and locally-conserved fields evolving in time in high dimensions. Our results above suggest that the resulting dynamical phase diagrams can be very rich, with multiple dynamical phases competing across different fluctuation regimes. The cloning method reviewed in this paper seems the ideal tool to explore this interesting phenomenology. However, given the scaling of the number of clones with the system size (including its dimension) needed to properly sample a given rare event, it seems likely that the standard cloning method will be insufficient to explore these events, and future work must focus on the application of the recent improvements of this method to compute rare event statistics in more realistic systems.

\begin{acknowledgments}
We thank Robert Jack for insightful discussions. The research leading to these results has received funding from the 
Spanish Ministry MINECO project FIS2017-84256-P. Additionally, this study has been partially financed by the Consejer\'ia de Conocimiento, Investigaci\'on y Universidad, Junta de Andaluc\'ia and European Regional Development Fund (ERDF), ref. SOMM17/6105/UGR.C.P.E. acknowledges the funding received from the European Union's Horizon 2020 research and innovation programme under the Marie Sklodowska-Curie Cofund Programme Athenea3I Grant Agreement No. 754446. We are also grateful for the computational resources and assistance provided by PROTEUS, the super-computing center of Institute Carlos I in Granada, Spain. 
\end{acknowledgments}

\end{document}